\documentclass[12pt, letterpaper, twoside]{article}
\usepackage{amsmath,float}
\usepackage{lineno,hyperref,amsmath,float,graphicx}

\begin{document}


\begin{center}\LARGE
Integrals of motion in time periodic Hamiltonian systems: The case of the Mathieu equation
\end{center}

\begin{center}\small A.C. Tzemos\footnote{ thanasistzemos@gmail.com} and G. Contopoulos\footnote{gcontop@academyofathens.gr}\\ \textit{Research Center for Astronomy and 
Applied Mathematics \\of the Academy of 
Athens\\ Soranou Efesiou 4, GR-11527 Athens, Greece}
\end{center}

\begin{abstract}
We present an algorithm for constructing analytically approximate  integrals of motion in
simple time periodic Hamiltonians of the form $H=H_0+
\varepsilon H_i$, where $\varepsilon$ is a perturbation parameter. We apply our algorithm in a Hamiltonian system whose dynamics is governed by the Mathieu equation and examine in detail the orbits and their stroboscopic invariant curves for different values of $\varepsilon$. We find the values of $\varepsilon_{crit}$ beyond  which  the orbits escape to infinity and construct integrals which are  expressed as  series in the perturbation $\varepsilon$ and converge up to $\varepsilon_{crit}$. In the absence of resonances the invariant curves are concentric ellipses which are approximated very well by our integrals.  Finally we construct an integral of motion which describes  the hyperbolic stroboscopic invariant curve of a resonant case.
\end{abstract}


\section{Introduction}
The detection of integrals of motion is an important task in the study of dynamical systems. The existence 
of integrals of motion  is  related to certain symmetries of the equations of motion of the system (through Noether's theorem) and gives a deep insight into their time evolution. Furthermore it simplifies significantly the calculations, since it decreases the independent variables of the system and can be used for the error control of the calculations.

The integrals of motion appear in two main categories: a) the exact integrals of motion and b) the approximate integrals of motion. In the case a) these integrals are analytical mathematical expressions of the variables of the system, that do not depend on time.
In the case b) the approximate integrals (also called formal integrals) are non-convergent series expansions, truncated at a certain order that remain approximately constant in time.

Formal integrals of motion have been proved to be very useful in Dynamical Astronomy and especially Galactic Dynamics. In particular, their construction in the case of autonomous dynamical systems has  already been considered in the past, even to high
orders using computer algebra (\cite{contopoulos1965resonance, gustavson1966constructing, giorgilli1979acorurerprocram, efthymiopoulos2005optimized}).

As regards time-periodic Hamiltonians, it was presented back in 1966 for the first time a general method for the calculation of formal integrals \cite{contopoulos1966adiabatic}. This method was applied  to some simple Hamiltonians of one and two degrees of freedom. However, due to the lack of computer algebra systems at that time, the calculations were made by hand and were limited to second order with respect o the perturbation parameter $\varepsilon$. 

In recent years there has been much interest in periodic in time Hamiltonians, especially in Russia. Most of this work dealt with stability problems in one or more degrees of freedom (e.g. Markeev \cite{markeyev1994third,markeev2004stability,
markeev2005multiple,MARKEYEV2006176,markeev2015birkhoff}, Kholostova \cite{kholostova1998non,KHOLOSTOVA2002529, kholostova2006resonant}, Bardin and Lanchares \cite{bardin2015stability}). More recently Bruno  (\cite{bruno2020normal,bruno2020normalization}) calculated normal forms in particular problems.

On the other hand some authors (Kandrup \cite{kandrup1998chaos, kandrup2003transient} and Terzi{\'c} and Kandrup \cite{terzic2004orbital}) have calculated orbit in time-periodic potentials, with emphasis on the generation of chaos.

Our interest in this problem was revived because of our need to compare the classical with the quantum mechanical results (Efthymiopoulos and Contopoulos \cite{efthymiopoulos2006chaos}).

In the present paper we exploit the power of the Maple Computer Algebra System in order to construct high order integrals of motion in simple time periodic Hamiltonian systems. As a first step towards this direction we apply our algorithm on a simple Hamiltonian system whose equations of motion
can be written  in a single   second order differential equation of the form:
\begin{align}
\frac{d^2x}{dt'^2}+\Big[a-2q\cos(2t')\Big]x=0.
\end{align}
Although this differential equation is linear in x and  there is no chaos in its orbits, it is very useful in Applied Mathematics. It is the well known Mathieu equation (ME) \cite{mclachlan1951theory, Richards} and  has applications in
various fields of physical sciences, such as Acoustics (e.g. in the study of an elliptical drum \cite{mathieu1868memoire}),  Quantum Mechanics (e.g. in the study of the quantum pendulum \cite{leibscher2009quantum}),  General Relativity (e.g. in the study of the solutions of wave equations in curved spaces \cite{birkandan2007examples}) and Quantum Chemistry (charged particle in a quadruple field \cite{fink2009physical}). Many applications of MEs can be found in \cite{ruby1996applications}.

As already known from the theory, different parameters in ME can lead to bounded or unbounded motion.  We make a detailed study of the orbits and their invariant curves on a stroboscopic surface of section by solving numerically in Python 3.7 the Hamilton equations. In the case of bounded orbits we  construct in Maple 2016 an integral of motion which is convergent for small perturbations and describes very well the invariant curves.  This integral becomes divergent and the forms of the orbits  change abruptly at the threshold of the escapes.

In section 2 of the present paper we give the Hamiltonian of our model
and describe our algorithm for the calculation of formal integrals of motion up to an arbitrary
order in the absence of resonances. Then 
in Section 3 we present our results in the case where
the perturbation parameter $\varepsilon$ is positive, by calculating both the orbits  and their stroboscopic sections (i.e. the distribution of their points after successive periods). We apply our algorithm and calculate the  integrals at successive orders of the perturbation parameter and show their convergence to the form of  the invariant curves on the stroboscopic sections. Then in Section 4 we examine the case of the negative values of the perturbation parameter and in Section 5 we study a formal integral in a  resonant case. Finally in Section 6 we summarize our results and draw our conclusion.

\section{Integral of the Mathieu equation}

In the special case of the Mathieu equation we have 
\begin{align}\label{ham}
H=H_0+ \varepsilon H_i=\frac{1}{2}(y^2+\omega_i^2x^2)-\varepsilon x^2\cos(\omega t)
\end{align}
The corresponding equations of motion are:
\begin{align}
\frac{dx}{dt}=y, \quad \frac{dy}{dt}=\frac{d^2x}{dt^2}={-\Big[\omega_1^2-2\varepsilon\cos(\omega t)\Big]x}
\end{align}
The second equation takes the form of the Mathieu equation if we set $\omega t=2t', a=4\omega_1^2/\omega^2$ and $q=\frac{4\varepsilon}{\omega^2}$. In particular if $\omega=2$ we have $t=t', a=\omega_1^2$ and $q=\varepsilon$. There are  resonance conditions when $a=1,4,9,16\dots$. For $\omega=2$ resonances appear if $\omega_1=1,2,3,\dots$
An integral of motion 
\begin{align}
\Phi=\Phi_0+\varepsilon\Phi_1+\varepsilon^2\Phi_2+\dots+\varepsilon^s\Phi_S+\dots
\end{align}
must satisfy the equation 
\begin{align}
\frac{d\Phi}{dt}=\frac{\partial \Phi}{\partial t}+[\Phi,H]=0,
\end{align}
where $[\Phi, H]$ is the Poisson bracket
\begin{align}
[\Phi,H]\equiv \frac{\partial \Phi}{\partial x}\frac{\partial H}{\partial y}-\frac{\partial \Phi}{\partial y}\frac{\partial H}{\partial x}.
\end{align}
We apply Eq.(3) to the terms of successive orders in $\varepsilon$ and find
\begin{align}\label{pde}
\frac{\partial\Phi_{s+1}}{\partial t}+[\Phi_{s+1},H_0]-K_s=0,
\end{align}
where 
\begin{align}
 {K_s=-[\Phi_s,H_1]}
\end{align}
If we set $\Phi_0=H_0$ we can calculate successively the terms of various orders of $\Phi$ expressed in trigonometric terms 
of multiplicities of $\omega t$. From the characteristic curves of Eq.~\eqref{pde} we find
\begin{align}\label{char}
dt=\frac{dx}{y}=\frac{dy}{-\omega_1^2x}=\frac{d\Phi_{s+1}}{K_s}
\end{align}
The zero order solution \textbf{is} 
\begin{align}\label{zo}
x=\frac{\sqrt{2\Phi_0}}{\omega_1}\sin(\omega_1 t), \,\, y=\sqrt{2\Phi_0}\cos(\omega_1 t)
\end{align}
Consequently from Eqs.~\eqref{char} we get
\begin{align}
\Phi_{s+1}=\int_0^t K_{s}dt
\end{align}
where $K_S$ is expressed in trigonometric terms of multiples of $\omega t$  and $\omega_1 t$. After the integration we use again Eqs.~\eqref{zo} to express back the trigonometric terms of
$\Phi_{s+1}$ which contain $\omega_1t$ in terms of $y^2,x^2,xy$ multiplied
by trigonometric terms of multiples of $\omega t$. Consequently the integral $\Phi$, which is in practice truncated at some order, is $\Phi=\Phi(x,y,\omega t)$.

In particular  $\Phi_0=H_0$ and 
\begin{align}\Phi_1=\int_0^t\frac{\partial\Phi_0}{\partial y}\frac{\partial H_1}{\partial x}dt=-2\int_0^txy\cos(\omega t)dt
\end{align}
The integral $\Phi$ is a series in $\varepsilon$ and of second order 
in $x,y$. Up to second order in $\varepsilon$
it is:
\begin{align}
\Phi=\frac{1}{2}\Big(C_x{\omega_{1}}^{2}{x}^{2}+C_y{y}^{2}+C_{xy}xy\Big),
\end{align} 
where
\begin{align}
\nonumber C_x=1&+{\frac {4\varepsilon \left( \cos \left( \omega t \right) +1
 \right) }{{\omega}^{2}-4{\omega_{1}}^{2}}}\\&\nonumber+{\frac {{\varepsilon}^{2}\left( 16 \left( {\omega}^{2}-{\omega
_{1}}^{2} \right) \cos \left( \omega t \right) + \left( 4{\omega_{1
}}^{2}+7{\omega}^{2}-{\frac {2{\omega}^{4}}{{\omega_{1}}^{2}}}
 \right) \cos \left( 2\omega t \right) +12{\omega_{1}}^{2}-7{
\omega}^{2}+{\frac {4{\omega}^{4}}{{\omega_{1}}^{2}}} \right)}{
 \left( {\omega}^{2}-4{\omega_{1}}^{2} \right) ^{2} \left( {\omega}^
{2}-{\omega_{1}}^{2} \right) } }\\&+\dots
\end{align}
\begin{align}
\nonumber  C_y=1&-{\frac {4\varepsilon \left( \cos \left( \omega t \right) -1
 \right) }{{\omega}^{2}-4{\omega_{1}}^{2}}}\\&+{\frac {{\varepsilon}^{2}
 \Big( -16 \left( {\omega}^{2}-{\omega_{1}}^{2} \right) \cos
 \left( \omega t \right) +3 \left( {\omega}^{2}-4{\omega_{1}}^{2}
 \right) \cos \left( 2\omega t \right) +13{\omega}^{2}-4{\omega
_{1}}^{2} \Big) }{ \left( {\omega}^{2}-4{\omega_{1}}^{2} \right) ^
{2} \left( {\omega}^{2}-{\omega_{1}}^{2} \right) }}+\dots
\end{align}
and
\begin{align}
C_{xy}={\frac {-4\varepsilon\omega \sin \left( \omega t \right) }{{\omega}
^{2}-4{\omega_{1}}^{2}}}+{\frac {{\varepsilon}^{2}\omega \Big( -16
 \left( {\omega}^{2}-{\omega_{1}}^{2} \right) \sin \left( \omega t
 \right) +6 \left( {\omega}^{2}-4{\omega_{1}}^{2} \right) \sin
 \left( 2 \omega t \right)  \Big) }{ \left( {\omega}^{2}-4{
\omega_{1}}^{2} \right) ^{2} \left( {\omega}^{2}-{\omega_{1}}^{2}
 \right) }}+\dots
\end{align}

We notice that the first order terms in $\varepsilon$ contain $\cos(\omega t), \sin(\omega t)$ and a constant, the second order terms contain also $\cos(2\omega t)$ and $\sin(2\omega t)$ and so on.
The constant terms are due to the fact that the integral starts at $t=0$.

\begin{figure}[H]
\centering
\includegraphics[scale=0.35]{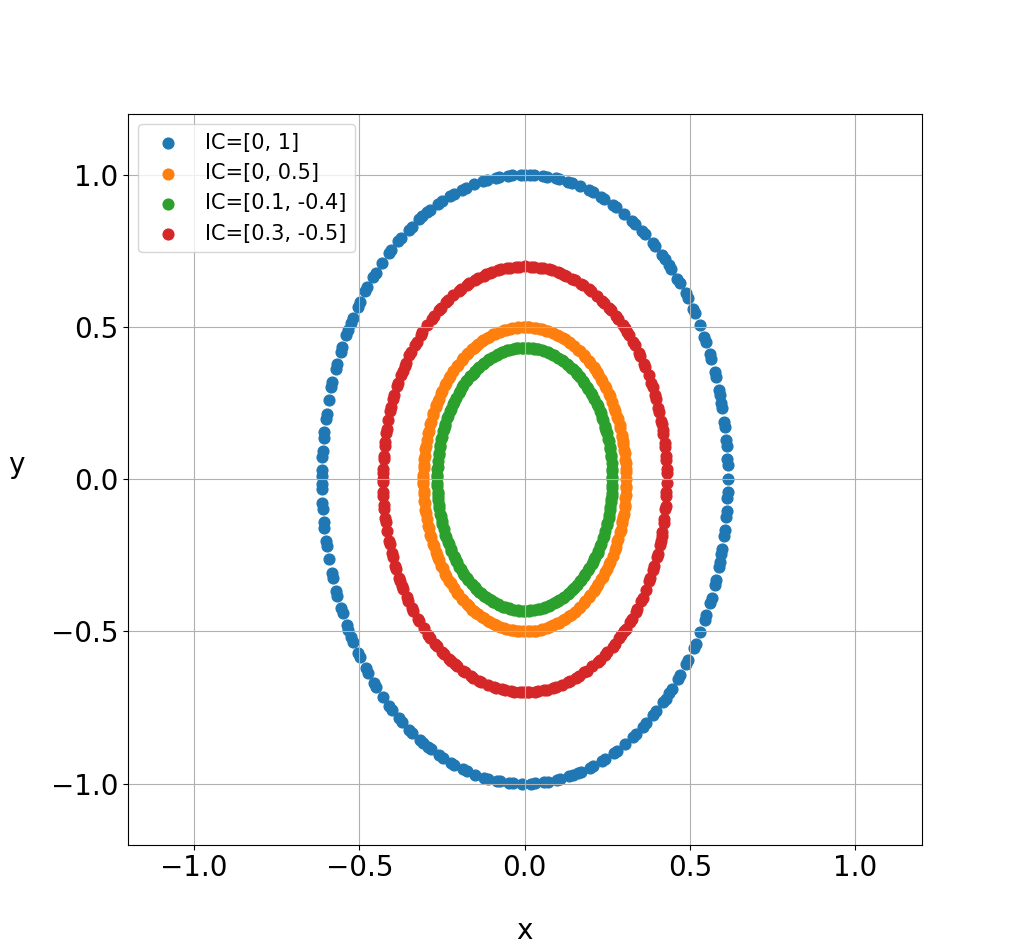}
\caption{The invariant curves (ellipses) of the stroboscopic surface of section in the case $\omega=2, \omega_1=0.9, \varepsilon=0.1$ and various initial conditions (IC) for the first 200 periods. The various concentric ellipses are similar to each other.}\label{strobes_sinthikes}
\end{figure}

In general the terms of order $\varepsilon^m$ contain cosines of multiples of $\omega t$ up to 
order $m\omega t$ together with constant terms in the 
coefficients of $x^2$ and $y^2$, while they contain only sines up to order $m\omega t$ in the coefficient of $xy$. The denominators 
contain factors of the form $(\omega^2-4\omega_1^2), (\omega^2-\omega_1^2),(\omega^2-9\omega_1^2),\dots (\omega^2-m^2\omega_1^2)$.

Using our program for the construction of this integral  
we have calculated its terms up to order 28.
For $\varepsilon$ not large the integral $\Phi$ represents an 
ellipse  of the form
\begin{align}
\Phi=Ax^2+By^2+2Dxy
\end{align}

If $t=2k\pi/\omega=kT$ we have
\begin{align}
A=\frac{\omega_i^2}{2}\Big[1+\frac{8\varepsilon}{\omega^2-4\omega_1^2}+\frac{2\varepsilon^2\omega^2(\omega^2+8\omega_1^2)+\dots}{\omega_1^2(\omega^2-4\omega_1^2)^2(\omega^2-\omega_1^2)}\Big],\quad B=\frac{1}{2}, D=0\label{eqA}
\end{align}
This represents an ellipse  passing through the initial point $(x_0=0, y_0=1)$ with semiaxes $a=\sqrt{\frac{\Phi}{A}},
b=\sqrt{\frac{\Phi}{B}}$. The points $(x,y)$ of an orbit on a stroboscopic 
Poincar\'{e} surface of section $t=kT$ lie on this ellipse, if the series giving $A$ converges.

We have calculated several orbits and verified that if 
$\varepsilon$ is small and $\omega$ is not close to a resonance 
$\omega=\omega_i,2\omega_i,3\omega_i...$ the points of the orbits 
on the stroboscopic surface of sections lie on such an ellipse. 
Figure \ref{strobes_sinthikes} represents a set of such ellipses for $\varepsilon=0.1, 
\omega=2$ and $\omega_1=0.9$. In fact for various initial conditions we 
have similar concentric ellipses.

\section{Applications for $\varepsilon>0$}

\begin{figure}[H]
\centering
\includegraphics[scale=0.35]{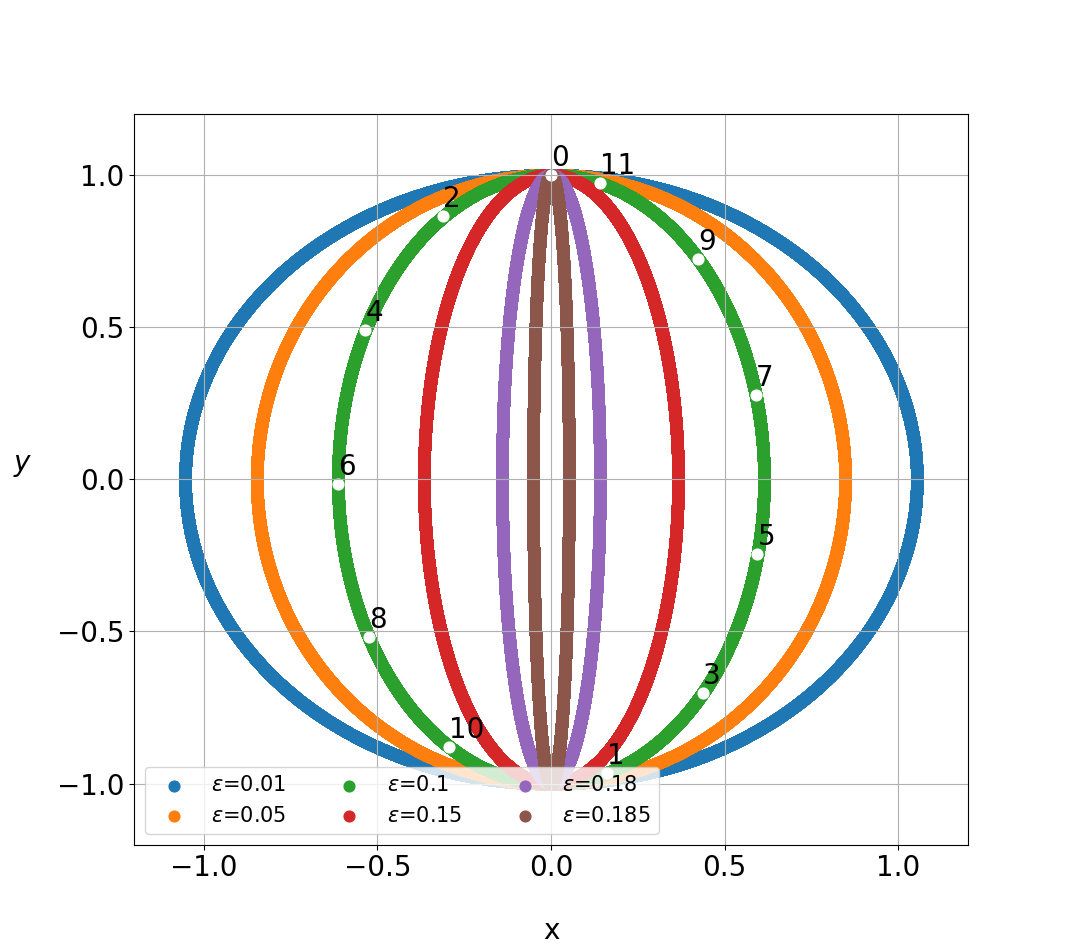}
\caption{Invariant curves on the stroboscopic surface of section in the case with $\omega=2, \omega_1=0.9$ for various values of $\varepsilon$ and the same initial conditions ($x_0=0, y_0=1$). In particular we have given the successive points $1, 2,\dots, 11$ covering clockwise the invariant curve for $\varepsilon=0.1$.}\label{strobes_thetika_epsilon}
\end{figure}

If we fix $\omega=2$ and $\omega_1=0.9$, and initial condition $x_0=0, y_0=1$  we have
 $\Phi=1/2$. Then the semiaxes of the 
ellipses are $a=\frac{1}{\sqrt{2A}}, b=1$.
In Fig.~\ref{strobes_thetika_epsilon} we present the stroboscopic surface of section for these parameters  for various values of $\varepsilon$. We observe that they are all  ellipses with one axis from $y=1$ to $y=-1$. The same holds for other non-resonant values of $\omega$ and $\omega_1$.

\begin{figure}[H]
\centering
\includegraphics[scale=0.3]{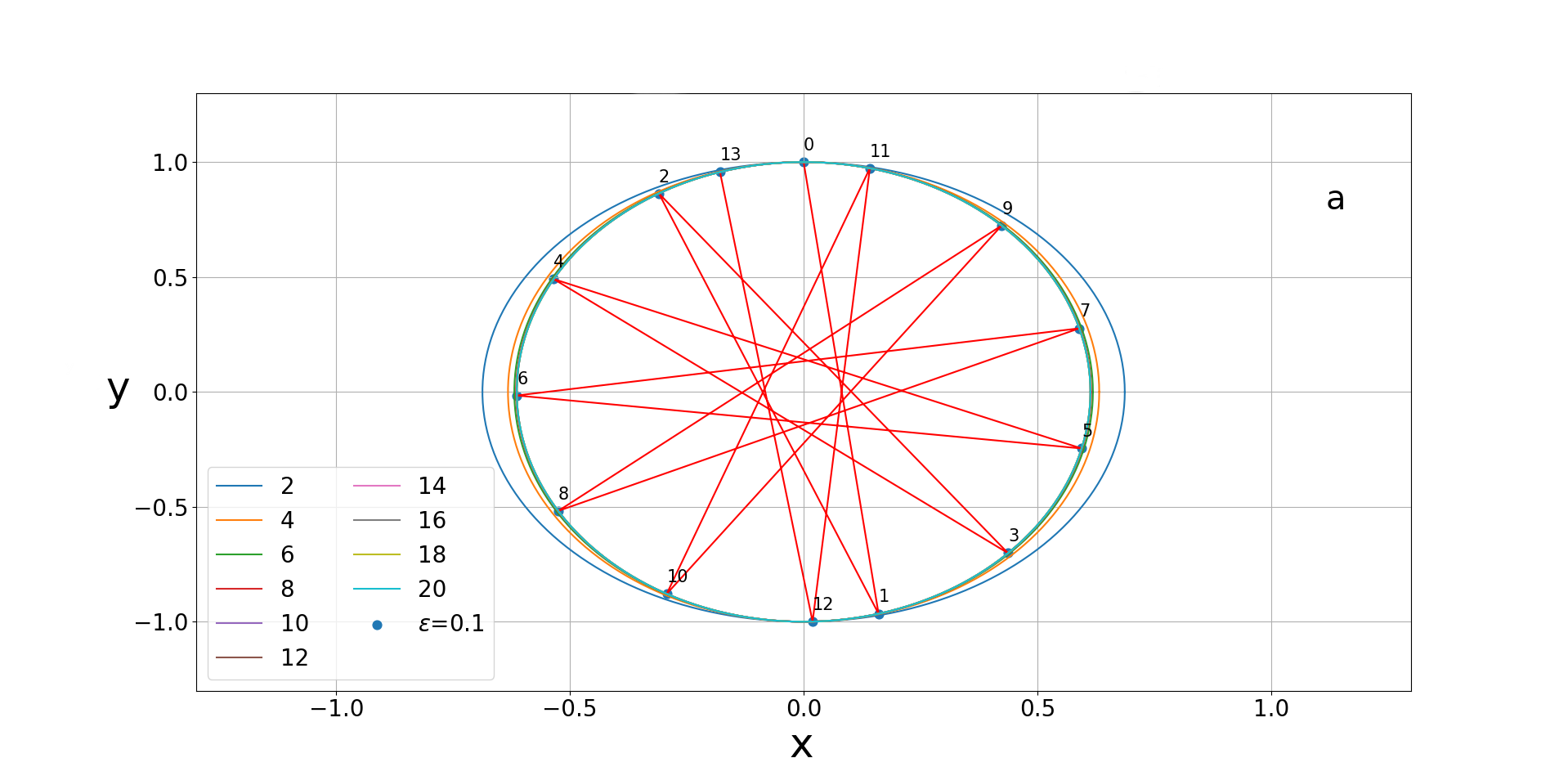}
\includegraphics[scale=0.3]{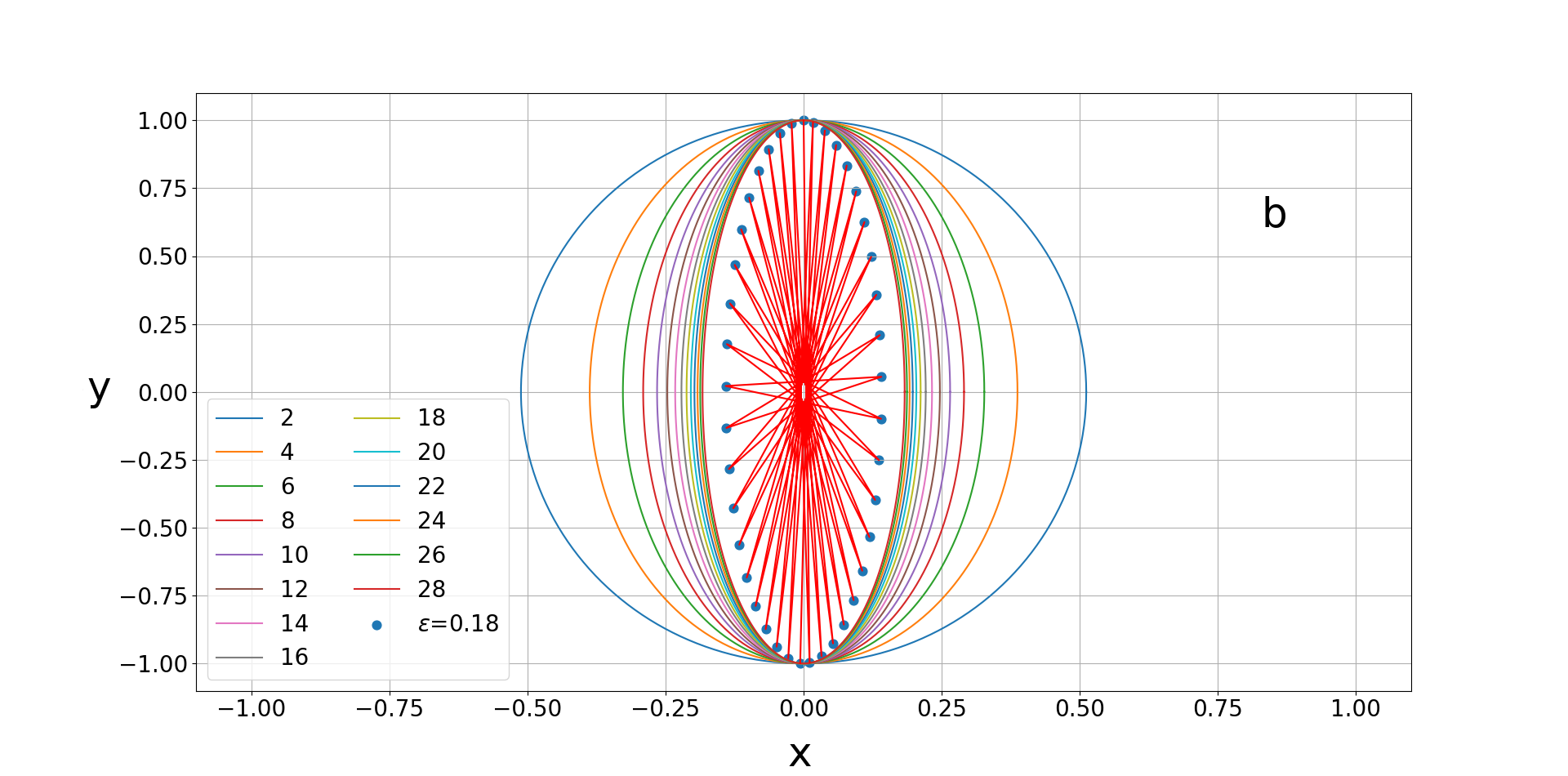}
\caption{Successive points of the orbits with $\omega=2, \omega_1=0.9$ and initial conditions ($x_0=0, y_0=1$) on the stroboscopic surface of section in the cases (a) $\varepsilon=0.1$ and (b) $\varepsilon=0.18$. The successive points are joined by straight lines. The points cover roughly for the first time the corresponding invariant curve. Their numbers are 13 in case (a) and 39 in case (b). The set of points is approached by the theoretical invariant curves   truncated at various successive orders as shown by colors.}\label{olokliromata_1}
\end{figure}

In Figs.~\ref{olokliromata_1}a,b we mark the successive points after times $T, 2T,3T\dots$ 
etc. for $\varepsilon=0.1$ and $\varepsilon=0.18$.  At the same time we draw the   ellipses found if we truncate the integral $\Phi$ after the terms of order $2, 4, 6$\dots in $\varepsilon$.
We see that the ellipse of order 2 is far from the points of the numerical 
solution. But as we increase the order of the truncation the ellipses 
approach gradually the invariant curve formed by the successive points. In the case 
$\varepsilon=0.1$ we have good convergence at order 6, in the case 
$\varepsilon=0.15$ good convergence is reached  at 
order 20, while in the case $\varepsilon=0.18$ we have not yet 
reached good convergence up to order 28. However these figures 
indicate that the integral $\Phi$ in fact converges all the way 
up to $\varepsilon=0.1857$, namely up to the critical value of 
$\varepsilon$ above which we have escapes to infinity.

Moreover we see that the angles between the lines joining the successive points decrease as $\varepsilon$ increases. In particular for $\varepsilon=0$ the successive points are on a circle on a plane ($\omega_1x,y$) and the successive angles are
$\Delta\phi=\frac{2\pi\omega_1}{\omega}=0.9\pi$. Thus the number of points required to cover roughly the invariant curve is about 11 (Fig.~\ref{olokliromata_1}a). In the case $\varepsilon=0.1$ this number is about 13, for $\varepsilon=0.15$ it is 17, for $\varepsilon=0.18$
it is  about 40  and for $\varepsilon=0.185$ it is  almost 110 points. The following iterations give additional points that fill more densely the ellipse provided by the integral $\Phi$.

\begin{figure}[H]
\centering
\includegraphics[scale=0.3]{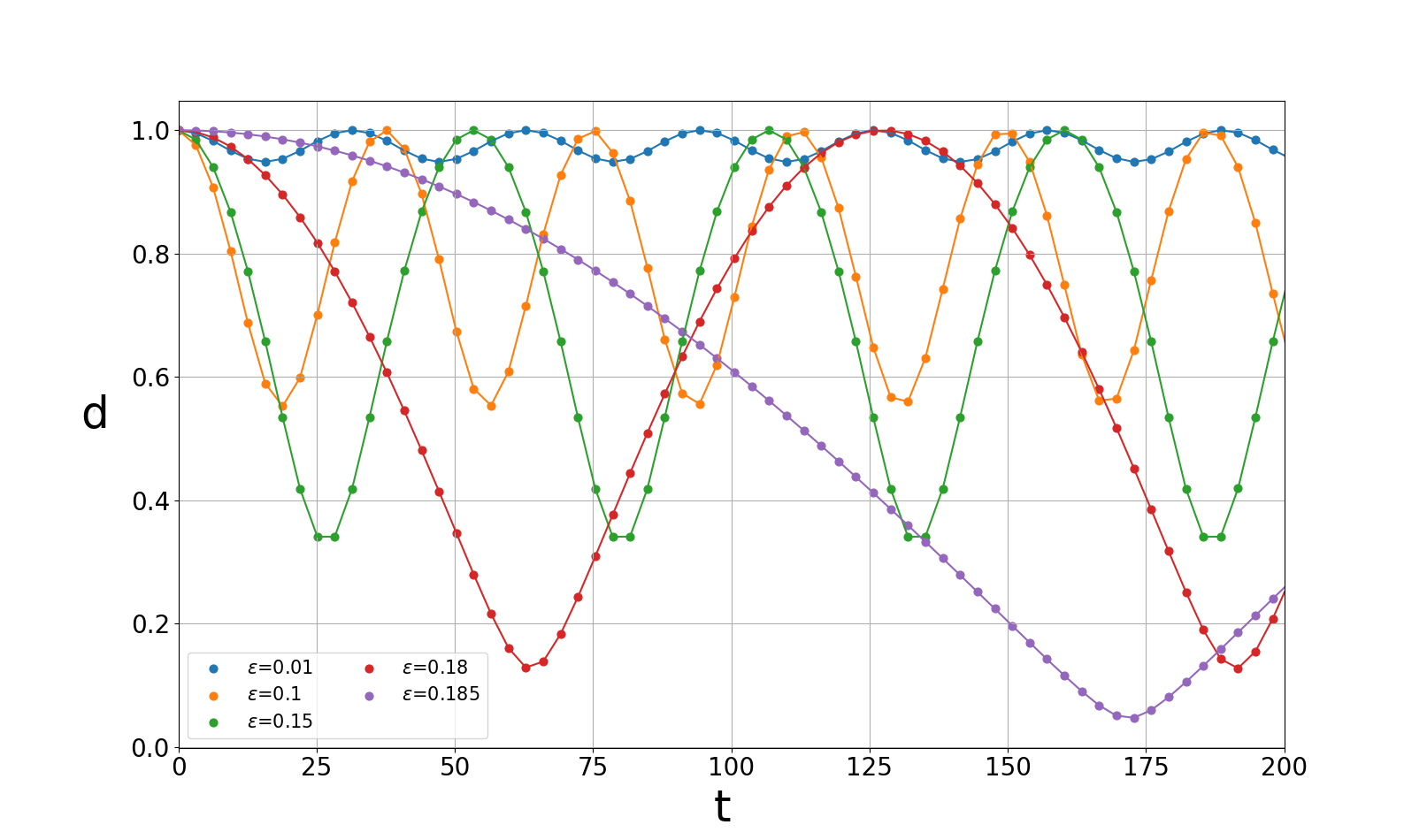}
\caption{The distances $d\equiv\sqrt{\omega_1^2x^2+y^2}$ between the origin $(0,0)$ and the successive points of the intersections of the orbits by the stroboscopic surface of section,  for various values of $\varepsilon$ up to $t=200$. We have $\omega=2, \omega_1=0.9, x_0=0, y_0=1$.}
\label{apostaseis}
\end{figure}

These numbers are found also 
by calculating the distances $d=\sqrt{\omega_1^2x^2+y^2}$ 
of the successive points on the stroboscopic surface of 
section as functions of time (Fig.~\ref{apostaseis}). All the orbits start
at $(x_0=0, y_0=1)$ with $d=1$ and decrease down to a minimum distance. Then they increase up to the distance $d=1$ and continue to decrease and increase. We see that for 
$\varepsilon=0.01$ one comes back to the maximum distance
after 11 points, for $\varepsilon=0.1$ after 13 points, 
and so on. Figure \ref{apostaseis} gives also the time required for the 
successive points to reach again the maximum distance. This
time increases considerably as $\varepsilon$ approaches the 
critical value $\varepsilon_{crit}=0.1857$.

The motion with initial conditions ($x_0=0,y_0=1$) for $\varepsilon<\varepsilon_{crit}$ starts by spiralling inwards until it reaches a minimum
distance  and then it spirals outwards and so on. Thus the orbit fills a ring as in Fig.5a for $\varepsilon=0.1$ and its corresponding invariant curve is an ellipse. As $\varepsilon$ increases the ring becomes broader and the empty hole near the center becomes smaller as in Fig. 5b for $\varepsilon=0.185$. Then the  invariant ellipse becomes very thin and in the limit $\varepsilon\to\varepsilon_{crit}$  the ellipse tends to the 
straight line from $y=1$ to $y=-1$ while the empty region  vanishes.

\begin{figure}[H]
\centering
\includegraphics[scale=0.25]{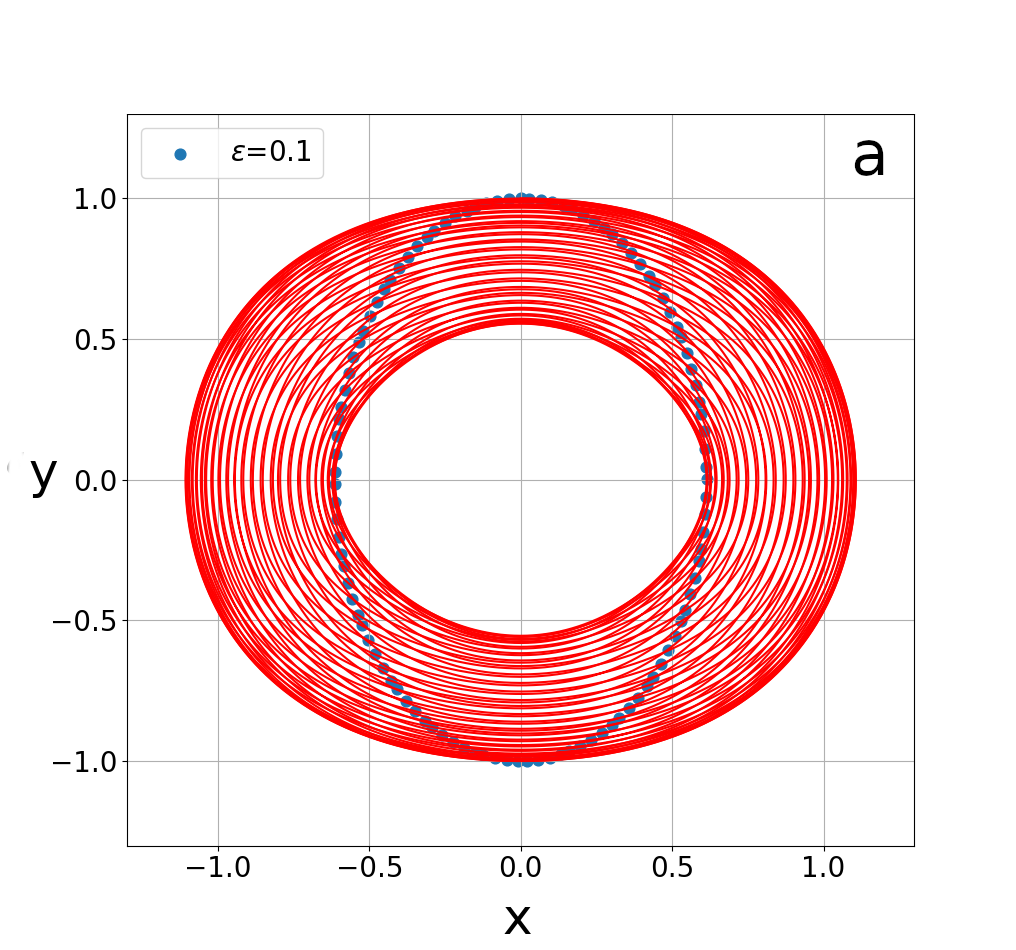}
\includegraphics[scale=0.25]{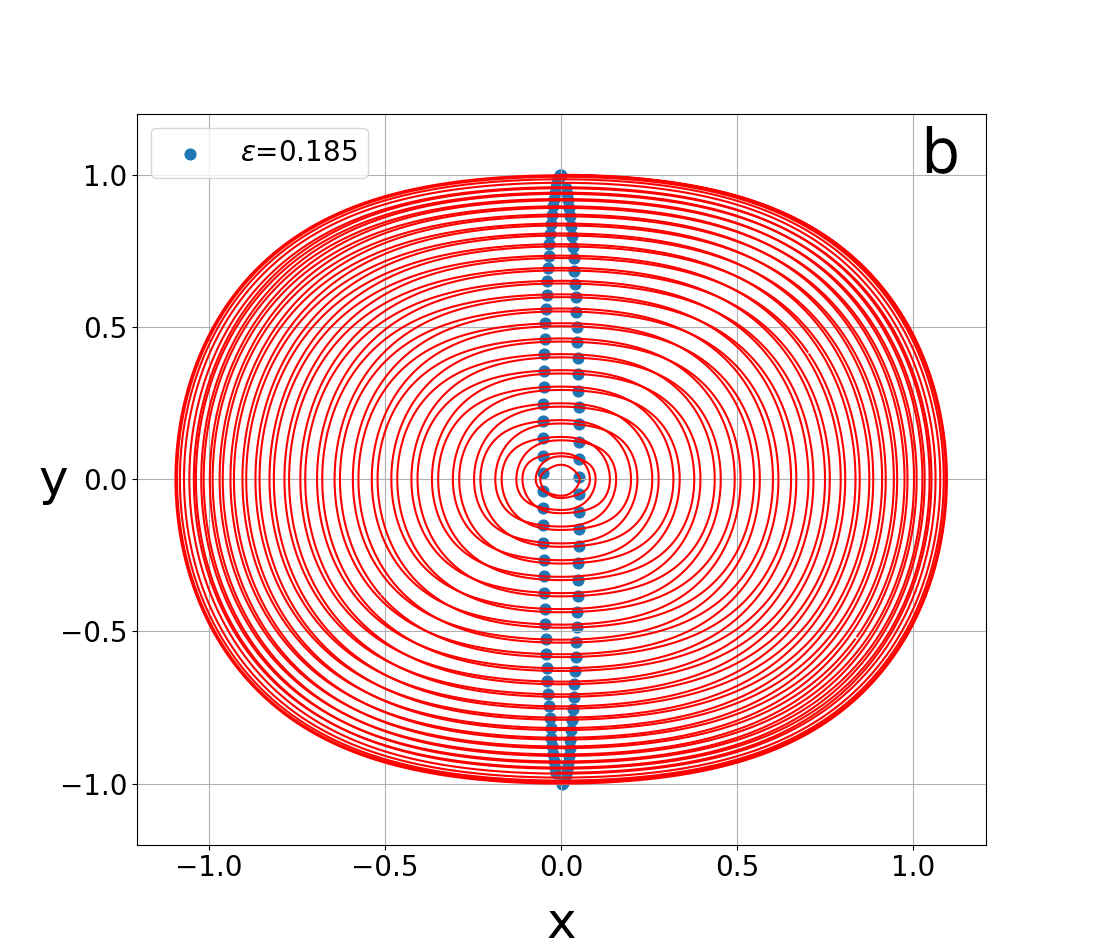}
\caption{Orbits in the case $\omega=2, \omega_1=0.9$ together with their intersections by the stroboscopic surface of section. (a) For $\varepsilon=0.1$ and 106 periods.  This orbit fills a ring inside the original arc of the orbit starting
at ($x_0=0, y_0=1$). (b) For $\varepsilon=0.185$ and 100 periods.
This orbit fills a ring leaving only a small hole around the center $(0,0)$.}
\label{strobe_epsilon_01}
\end{figure}

For particular values of $\varepsilon$ the orbit comes 
exactly to its initial point $x_0=0, y_0=1$ after a number
$n$ of periods i.e. we have a periodic orbit of period $nT$. Such is the case of Fig.~6 for $\varepsilon=0.15$, where we have a periodic orbit of period $17T$.  In this case the energy $E$ increases to a maximum, but then it returns to its initial value. Similar periodic orbits of other periods appear for other values of $\varepsilon$.

If $\varepsilon$ goes beyond a critical value about $\varepsilon_{crit}=0.1857$ the
orbits escape to infinity. In the present study we notice that the series $\Phi$ representing the  integral
of motion does not converge beyond the critical value of $\varepsilon$. E.g. for $\varepsilon=0.19$ (Fig.\ref{troxia_019}) the orbit starts close to an initial ellipse  but then makes a spiral outwards 
that extends to about $x=\pm15, y=\pm15$ after 50 periods. In Fig.~\ref{apostaseis_escape} we see that the logarithms of the distances for particular values of $\varepsilon$ beyond the critical value increase linearly in time. Therefore the distances of the escaping orbits increase exponentially in time and the increase is larger for larger $\varepsilon$. We know from the theory
that this motion extends to infinity.

\begin{figure}[H]
\centering
\includegraphics[scale=0.3]{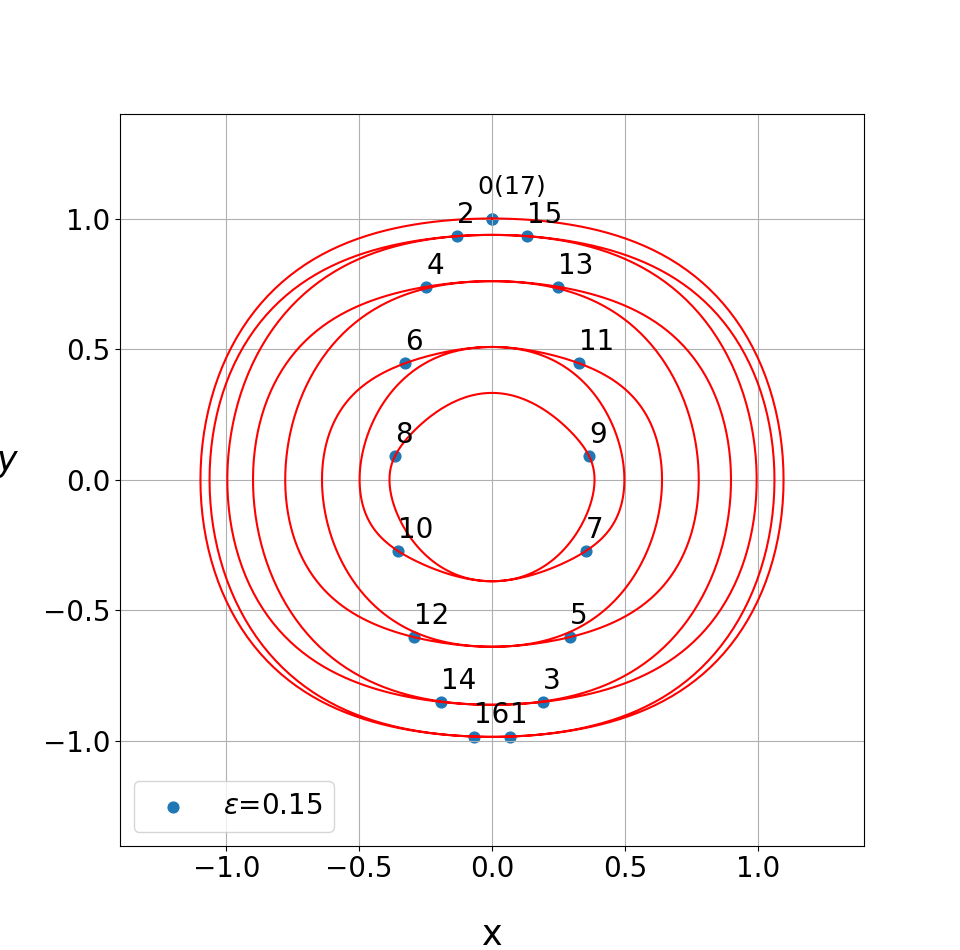}
\label{arithmimena6}
\caption{A periodic orbit of period $17T$ where $T=2\pi/\omega$, together with the 17 points of intersection with the stroboscopic surface of section.}
\end{figure}

\begin{figure}[H]
\centering
\includegraphics[scale=0.3]{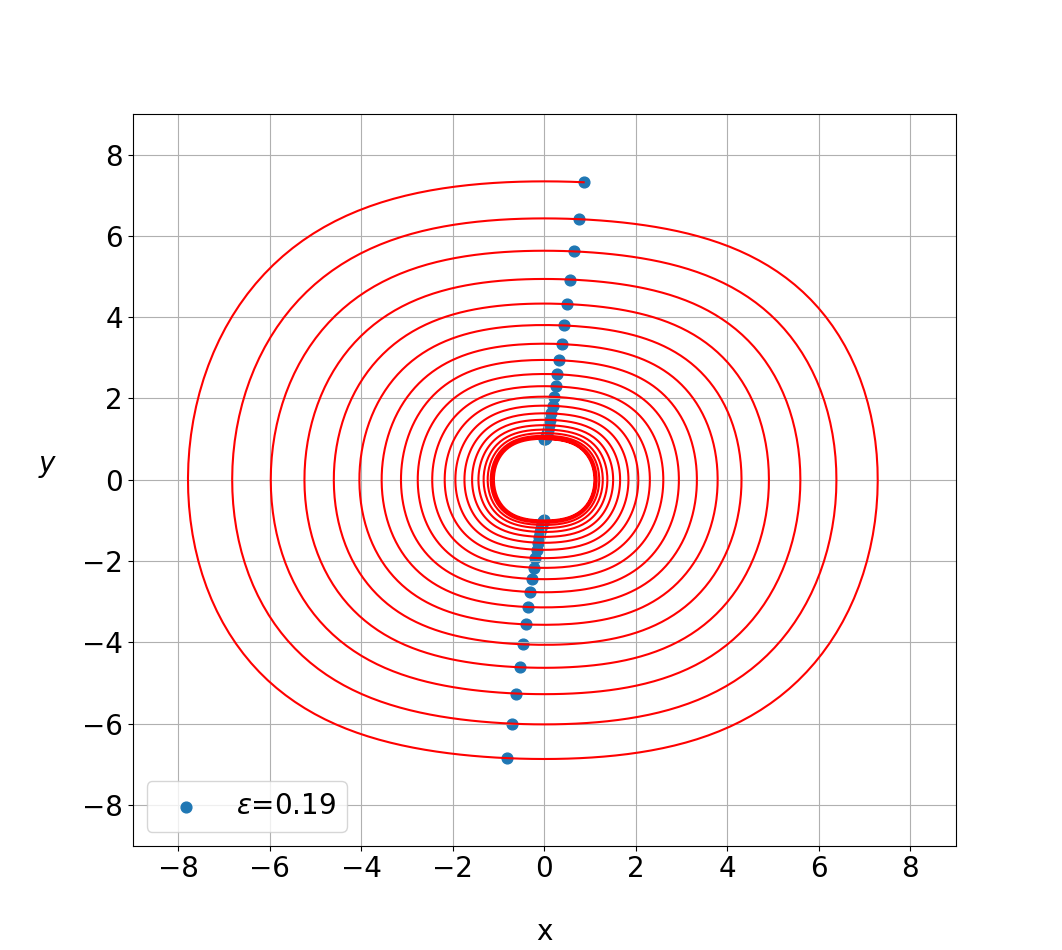}
\caption{An escaping orbit for ($\omega=2, \omega_1=0.9, \varepsilon=0.19$) starting at the point ($x_0=0, y_0=1$) for the first 40 periods, together with its intersections with the stroboscopic surface of section.}\label{troxia_019}
\end{figure}

\begin{figure}[H]
\centering
\includegraphics[scale=0.3]{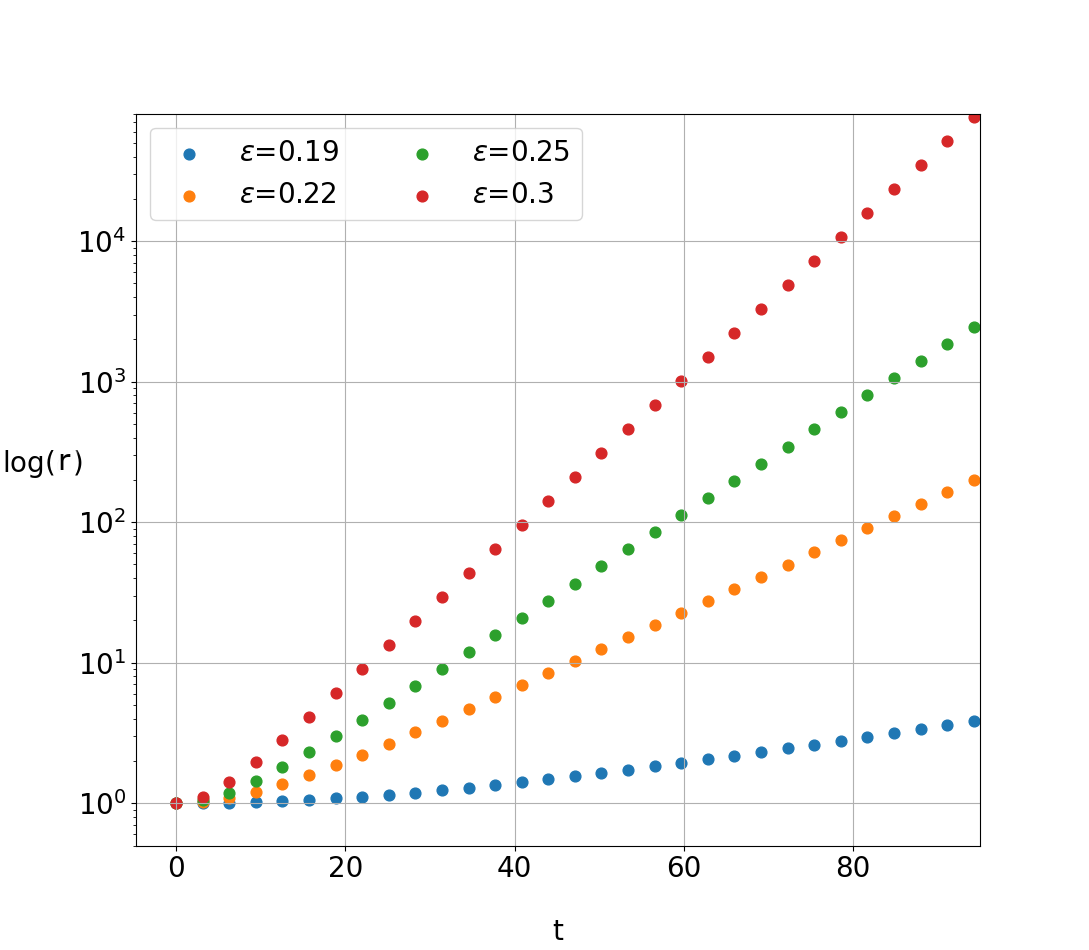}
\caption{The logarithms of the  distances  $r=\sqrt{x^2+y^2}$ of the successive intersections of escaping orbits by the stroboscopic surface of section for $\omega=2, \omega_1=0.9, x_0=0, y_0=1$ and  various values of $\varepsilon$ up to $t=30T$. In all these cases the logarithms of the distances increase linearly in time thus the distances icnrease exponentially in time. The distances are longer for larger $\varepsilon$.}\label{apostaseis_escape}
\end{figure}

An interesting aspect of our problem is found 
if we work in the so called extended phase space where our system becomes
conservative. Namely we extend the phase space to include the ordinary time as a canonical variable and its conjugate momentum $E$ which is minus the energy. Then the new Hamiltonian \begin{align}
\bar{H}(x,y,E,t)=H(x,y,t)+E
\end{align}
is conservative with respect to a fictitious time variable $\tau=t$ and the Hamilton equations read:
\begin{align}\label{eqs}
&\nonumber\frac{dx}{dt}=\frac{\partial H}{\partial y}, \quad
\nonumber\frac{dy}{dt}=-\frac{\partial H}{\partial x}\\&
\frac{d\tau}{dt}=\frac{\partial \bar{H}}{\partial E}=1, \quad
\frac{dE}{dt}=-\frac{\partial \bar{H}}{\partial t}=-\frac{\partial {H}}{\partial t}
\end{align}

\begin{figure}[H]
\centering
\includegraphics[height=5cm,width=6cm]{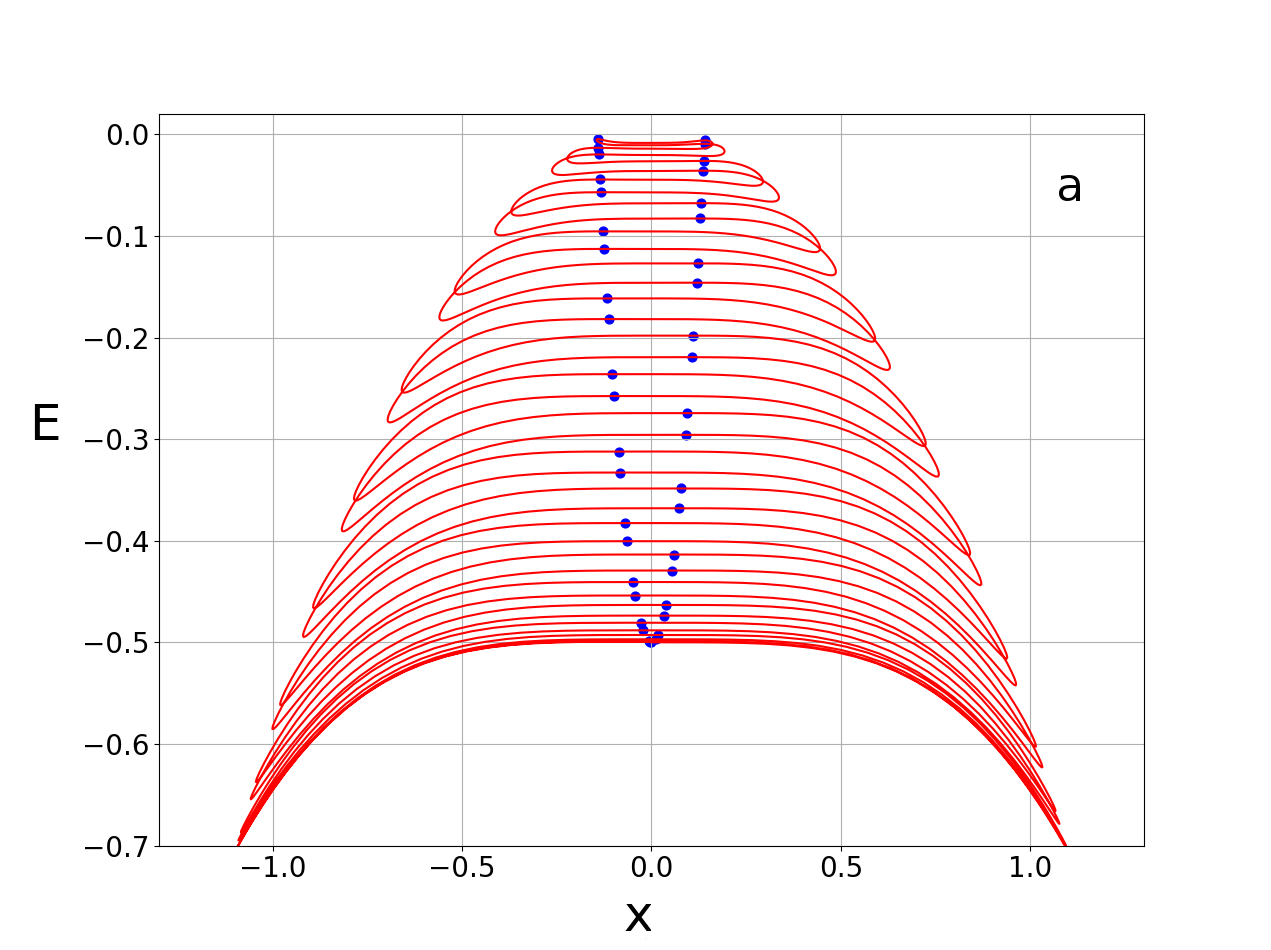}
\includegraphics[height=5cm,width=6cm]{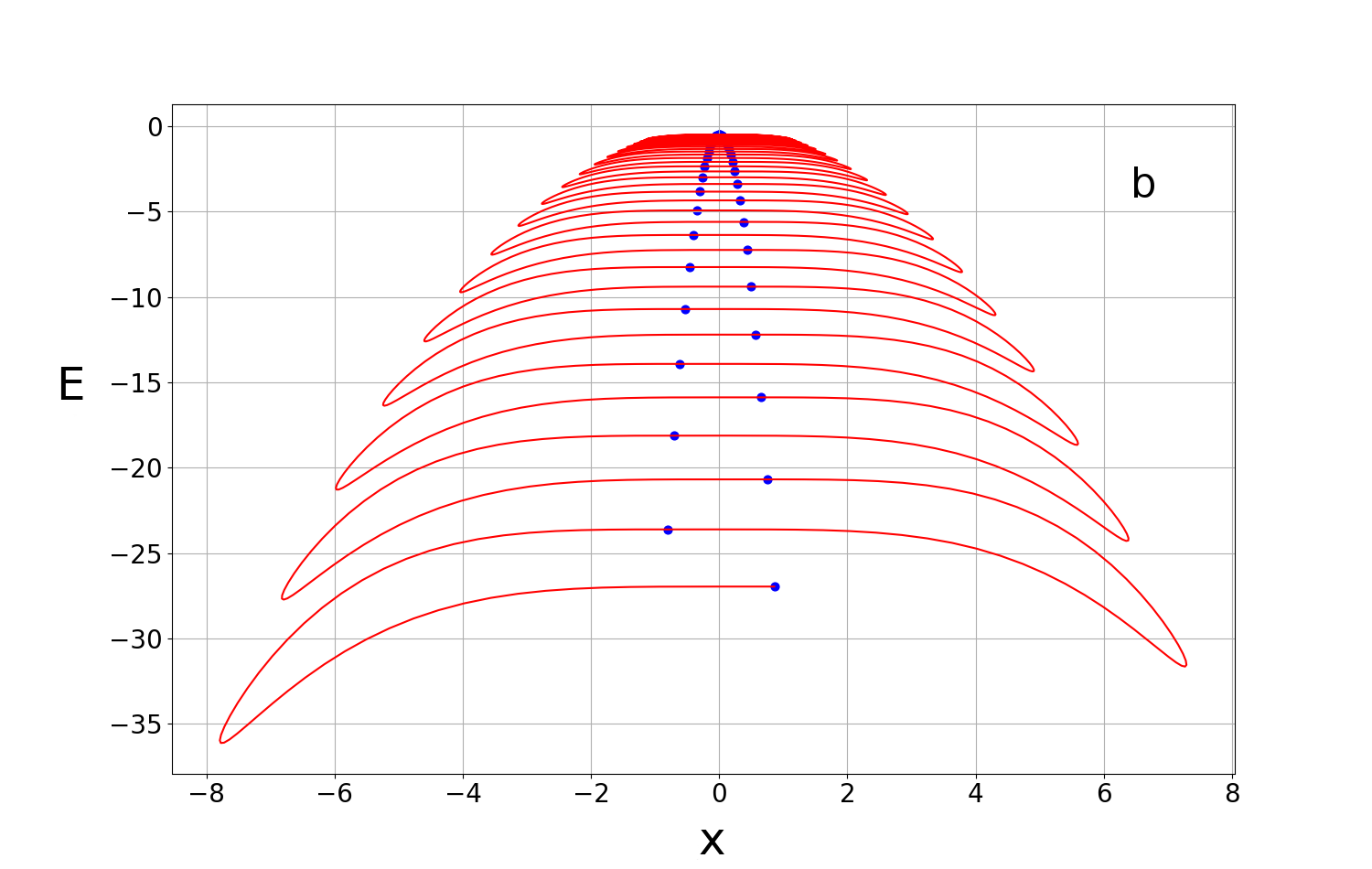}
\caption{Stroboscopic surfaces of section (blue points).  The trajectory of $E$ as a function of $x$ for (a) $\varepsilon=0.18$ up to 41 periods and (b) for $\varepsilon=0.19$ up to 40 periods. We observe that in the case (a) the trajectory is confined and the blue points  oscillate in a certain range of energies, while in the case (b) the trajectory escapes to $-\infty$ and the blue points also tend to $-\infty$. In the extended space, the total energy $\bar{H}=H+E$ is conserved and equal to 0 for our parameters.}
\label{elleipseis_arnitika_epsilon}
\end{figure}

Then we can calculate the value of the energy $E$ as a 
function of time using the last Eq.~\eqref{eqs} and we find that the value of $\bar{H}$ is very close to zero (in fact it is of  order $\mathcal{O}(10^{-8})$ which is the accuracy of our calculations). In the cases $0<\varepsilon<\varepsilon_{crit}$  for $x_0=0, y_0=1$
the value of $E$ starts at $E=-0.5$ for $t=0$, reaches a maximum near $E=0$ and then oscillates between this maximum  and a minimum. The values of $x$ and $E$ of the stroboscopic section are given by blue dots (Fig.~\ref{elleipseis_arnitika_epsilon}a). These dots mark  the points of the invariant curve on the stroboscopic section. The first 40 points are successively on the right and on the left part of the hyperbolic-like invariant curve which is directed upwards. Further points fill this curve densely. On the other hand for $\varepsilon>\varepsilon_{crit}$ the value of $E$ decreases 
continuously on the average (Fig.~\ref{elleipseis_arnitika_epsilon}b) and tends to
$-\infty$. The dots are again on a hyperbola-like curve, but this time the curve is directed downwards and has no limit.
This difference allows us to find with great accuracy the transition value $\varepsilon_{crit}$ for the escapes. At $\varepsilon=\varepsilon_{crit}\simeq0.1857848626$ the orbit is  periodic and it is  represented by the point 
$(x=x_0=0, y=y_0=1)$ on the stroboscopic surface of
section. We have calculated the monodromy matrix of this periodic orbit and found that its eigenvalues are all equal to one. Therefore the orbit is marginally unstable.

In the extended phase space we can calculate another integral 
\begin{align}\Psi=\Psi_0+\varepsilon\Psi_1+\varepsilon^2\Psi_2+\dots
\end{align}
if we set $\Psi_0=E$. Then we find
\begin{align}
\Psi_1=\frac{\left[    \left( 2{\omega_{1}}^{2}{-\omega}^{2}
 \right) {x}^{2}+2{y}^{2} \right] \cos \left( \omega\,t \right)+2\sin \left( \omega t \right) \omega xy-2
(\omega_1^2x^2+{y}^{2} )}{\omega^2-4\omega_1^2}
\end{align}
But then we notice that 
\begin{align}
\Phi_1+\Psi_1=- x^2\cos(\omega t)=H_1,
\end{align}
therefore 
\begin{align}
\Psi_1=H_1-\Phi_1
\end{align}
If we calculate now $\Psi_2$ this is equal to $-\Phi_2$ because $[H_1,H_1]=0$. In the same way we find $\Psi_3=-\Phi_3$ and so on. Thus
\begin{align}
\Phi+\Psi=\frac{1}{2}(\omega_1^2x^2+y^2)+E-\varepsilon x^2\cos(\omega t)=\bar{H}=0.
\end{align}
Consequently we have only two independent isolating integrals $\bar{H}$ and $\Phi$, thus the transformation of the system to an autonomous Hamiltonian system does not give any new results.

\begin{figure}[H]
\centering
\includegraphics[scale=0.28]{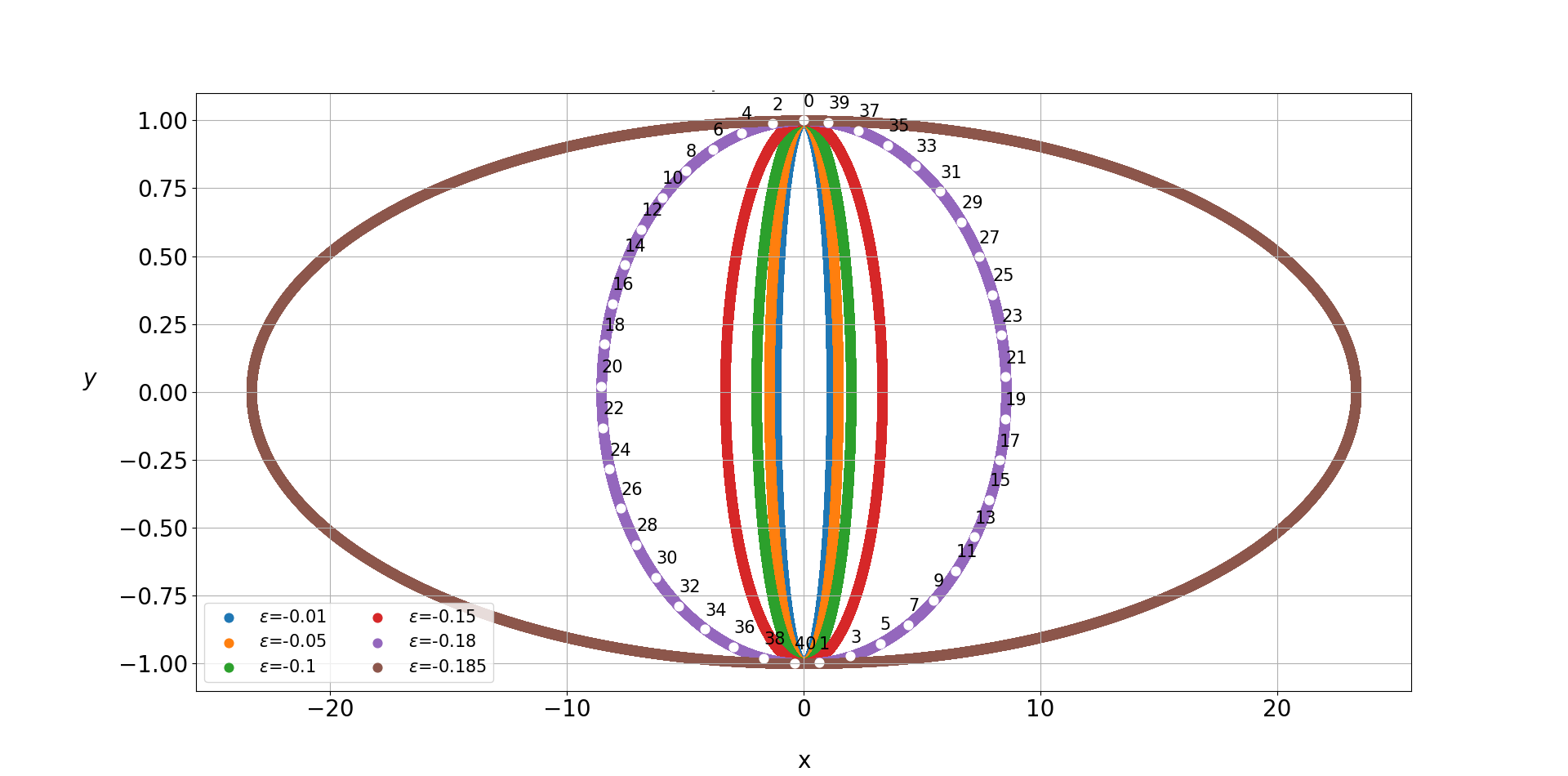}
\caption{Invariant curves on a stroboscopic surface of section in the case $\omega=2, \omega_1=0.9$ for $x_0=0,y_0=1$ and various negative values of $\varepsilon$. As $\varepsilon$ decreases and approaches the critical value $\varepsilon_{crit}\simeq -0.1857$ the size of the curve along $x$  increases and tends to $\infty$. In the case $\varepsilon=-0.18$ we give also the successive points corresponding to successive periods.}
\label{elleipseis_arnitika_epsilon2}
\end{figure}

\begin{figure}[H]
\centering
\includegraphics[scale=0.25]{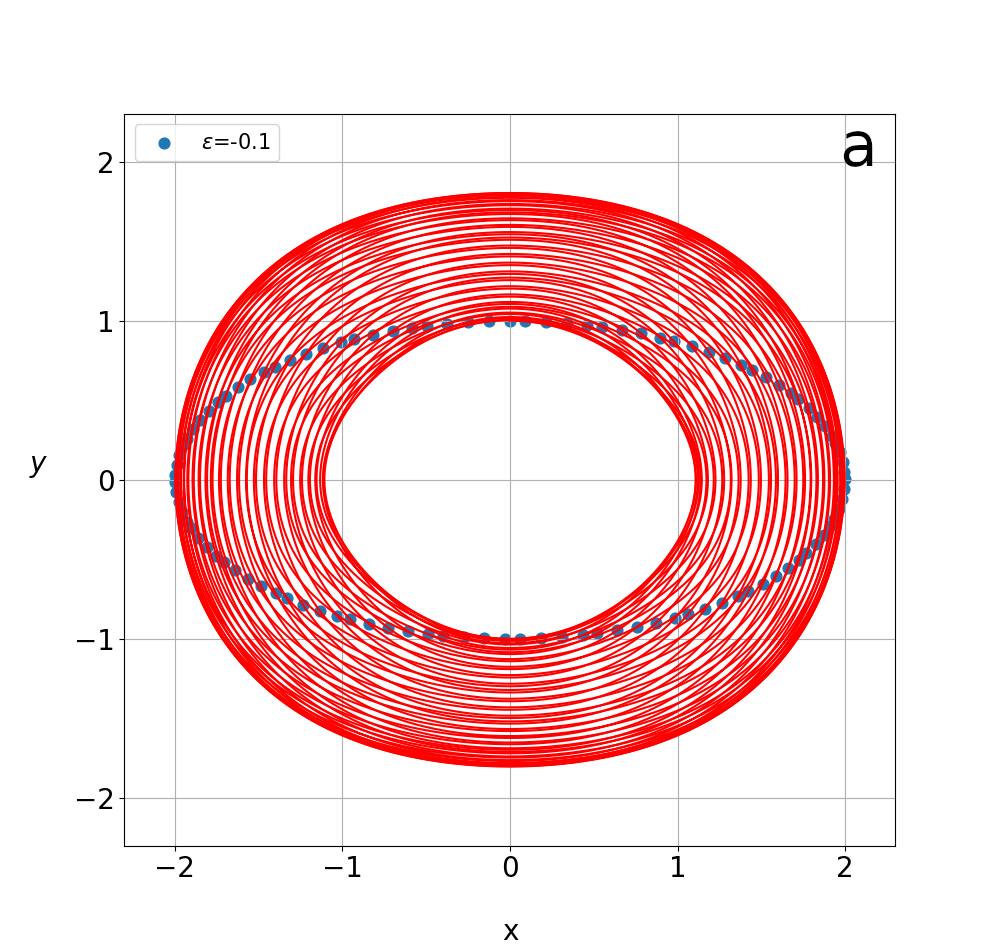}
\includegraphics[scale=0.25]{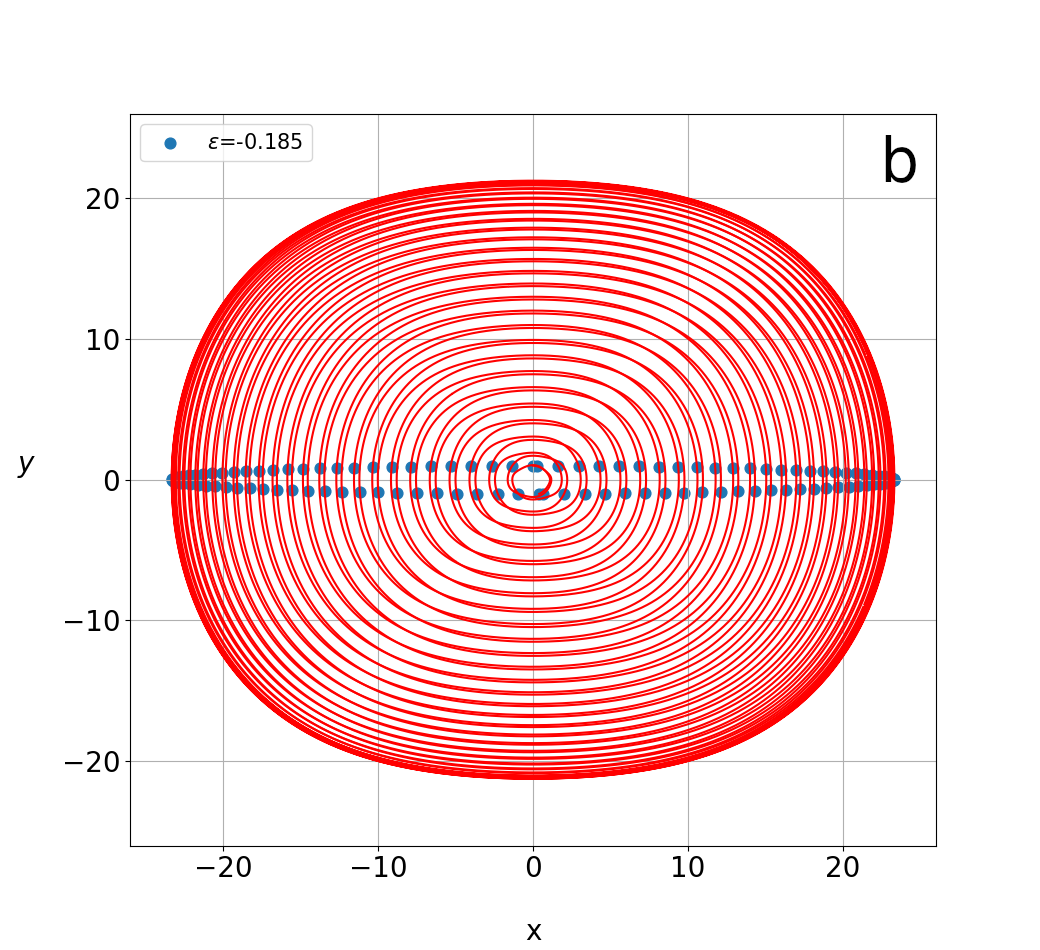}
\caption{Orbits in the case $\omega=2, \omega_1=0.9$ (a) For $\varepsilon=-0.1$ and 106 periods. The orbit fills a ring outside the original arc  of the orbit starting at ($x_0=0, y_0=1$) (b) For $\varepsilon=-0.185$ and 110 periods. We observe that the orbit extends to much larger distances than in the case (a).}
\label{arnitika_epsilon}
\end{figure}

\section{Cases with $\varepsilon<0$}
If $\varepsilon<0$ we have again ellipses on the stroboscopic Poincar\'{e}
surface of section. However these ellipses are outside the ellipse for 
$\varepsilon=0$ (Fig.~\ref{elleipseis_arnitika_epsilon2}). The critical value of $\varepsilon$ is $\varepsilon_{crit}=
-0.1857$, symmetric to the critical value $\varepsilon_{crit}=0.1857$ for
positive $\varepsilon$. The stroboscopic ellipses become longer along
the x-axis as $\varepsilon$ increases and
they tend to infinity as $\varepsilon$ tends to the critical value.

\begin{figure}[H]
\centering
\includegraphics[scale=0.3]{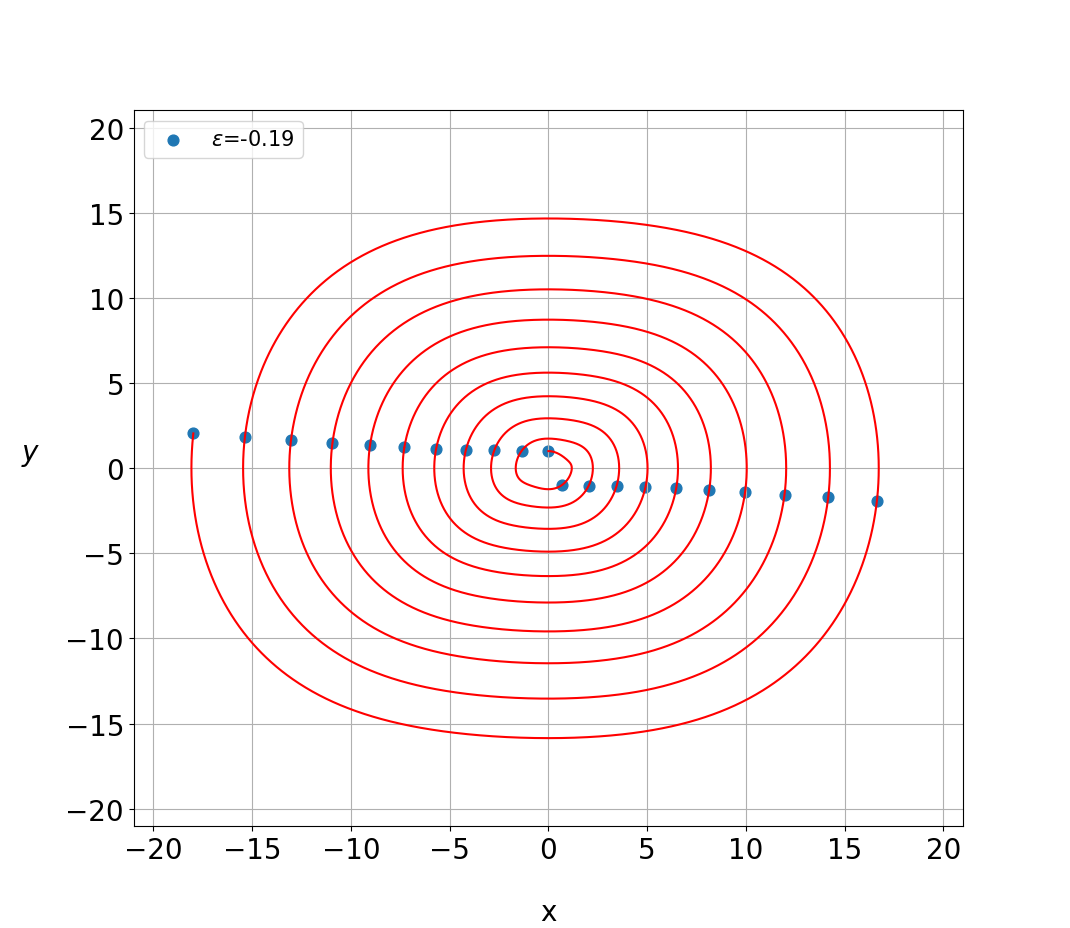}
\caption{An escaping orbit for $\omega=2, \omega_1=0.9, x_0=0, y_0=1, \varepsilon=-0.19$ for the first 20 periods, together with its intersections with the stroboscopic surface of section.}\label{troxia_epsilon_meion_019}
\end{figure}

The orbits for $\varepsilon<0$ are   different from those of $\varepsilon>0$.
The orbits are outside the limiting ellipse $\omega_1^2x^2+y^2=1$ for
$\varepsilon=0$. The number of points on the stroboscopic ellipse required to cover once the whole ellipse
again increases as $|\varepsilon|$ increases. For example for $\varepsilon=-0.1$
this number is 39 (Fig.~\ref{elleipseis_arnitika_epsilon2}) and for $\varepsilon=-0.185$ it is $n=110$ (Figs~\ref{arnitika_epsilon}a,b).
The orbits for $\varepsilon<0$ extend to large distances both in $x$ and $y$
but with much smaller ellipticity than the corresponding stroboscopic ellipse.
Finally for $|\varepsilon|$ larger than the critical value $|\varepsilon_{crit}|
=0.1857$ the orbits  escape again to infinity. The form of such an orbit 
is shown in Fig.~\ref{troxia_epsilon_meion_019} for $\varepsilon=-0.19$.

If we change now the value of $\omega_1$ we find again ellipses for various 
values of $\varepsilon$ given by the integral of motion. However the critical
values of $\varepsilon$ are now different. We consider here two examples.

\begin{enumerate}
\item In the case $\omega_1=0.1$ we have the ellipses of Fig.~\ref{omega_1_01}. In this case the 
limiting ellipse $\omega_1^2x^2+y^2=1$ is just outside the ellipse for $\varepsilon=0.001$, i.e. it is much more elongated along the x-axis than in the case with $\omega_1=0.9$. 
Furthermore the successive points on the stroboscopic surface of section are much
closer to each other. The total number of points to cover roughly the  
ellipse once for $\varepsilon=0.1$ is $N\simeq 16$. As $\varepsilon$ increases this number  increases.
In this case the critical value of $\varepsilon$ is $\varepsilon_{crit}=0.89964$, i.e.
much larger that in the case $\omega_1=0.9$.
\item The second case is $\omega_1=1.1$, larger than the resonant
value $\omega_1=1$. In this case the critical value of $\varepsilon$ is about
$\varepsilon_{crit}=0.21598$. We have again ellipses
on the stroboscopic surface of section for $\varepsilon<\varepsilon_{crit}$
but these become larger as $\varepsilon$ increases  (Fig.~\ref{omega_1_comma1}a). The difference between the behaviour of the case $\omega_1=1.1$ and the cases $\omega_1=0.9$ and $\omega_1=0.1$ can be explained theoretically. In fact close to $\varepsilon=0$ the value of the coefficient $A$ of Eq.~\eqref{eqA} is  approximately
\begin{align}
A\simeq\frac{\omega_1^2}{2}\Big(1+\frac{8\varepsilon}{\omega^2-4\omega_1^2}\Big)
\end{align}
while $\Phi=1/2$ for $(x_0=0, y_0=1)$
hence 
\begin{align}
a\simeq \frac{1}{\omega_1}\Big(1-\frac{4\varepsilon}{\omega^2-4\omega_1^2}\Big)
\end{align} 
\begin{figure}[H]
\centering
\includegraphics[scale=0.31]{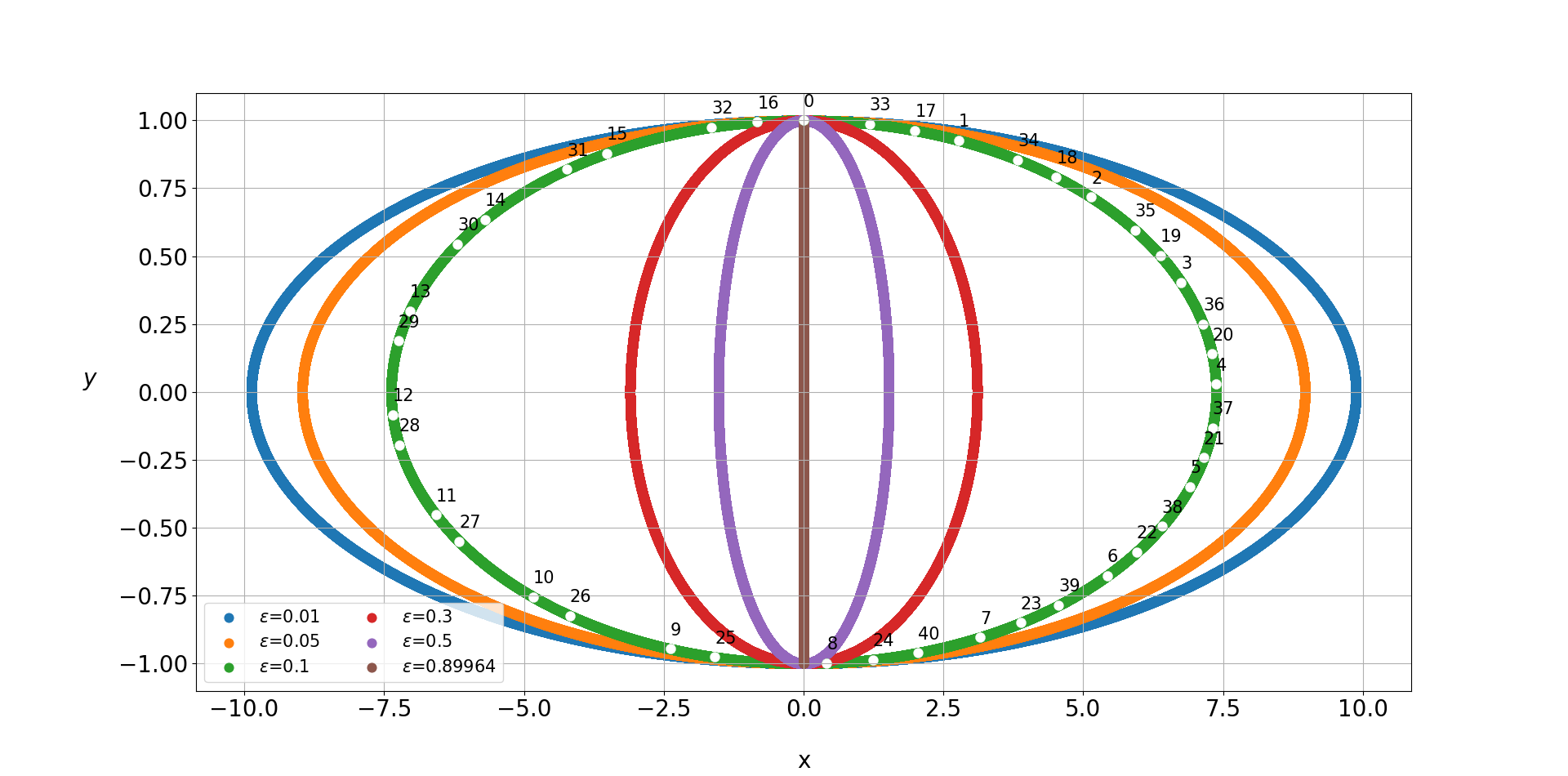}
\caption{Invariant curves (ellipses) on the stroboscopic surface of section on the case $\omega=2, \omega_1=0.1,x_0=0, y_0=1$, for various values of $\varepsilon>0$. As $\varepsilon$ increases the invariant curves shrink and for $\varepsilon=0.89964$ the curve is very elongated close to a critical value (almost a straight line). In the case $\varepsilon=0.1$ we give also the successive points.}\label{omega_1_01}
\end{figure}
Therefore for $\varepsilon>0$ and $(\omega^2-4\omega_1^2)>0$ we have $\varepsilon/(\omega^2-4\omega_1^2)>0$, hence 
$a<\frac{1}{\omega_1}$, i.e. the ellipses are inside the ellipse for $\varepsilon=0$, while for $\varepsilon<0$ and $(\omega^2-4\omega_1^2)>0$ we have $a>\frac{1}{\omega_1}$, i.e. the ellipses are outside the ellipse for $\varepsilon=0$. On the other hand for $(\omega^2-4\omega_1^2)<0$ the ellipses are larger for $\varepsilon>0$ and smaller for $\varepsilon<0$.

Thus
the case $\omega_1=1.1$ with $\varepsilon>0$  is similar to the case $\omega_1=0.9$ for negative $\varepsilon$.
On the other hand for $\varepsilon<0$ the ellipses become smaller (Fig.~\ref{omega_1_comma1}b)
for increasing $|\varepsilon|$ and  for $\varepsilon$ approaching the 
critical value $\varepsilon= -0.21598$ they tend to the x-axis.

\begin{figure}[H]
\centering
\includegraphics[height=9cm, width=12cm]{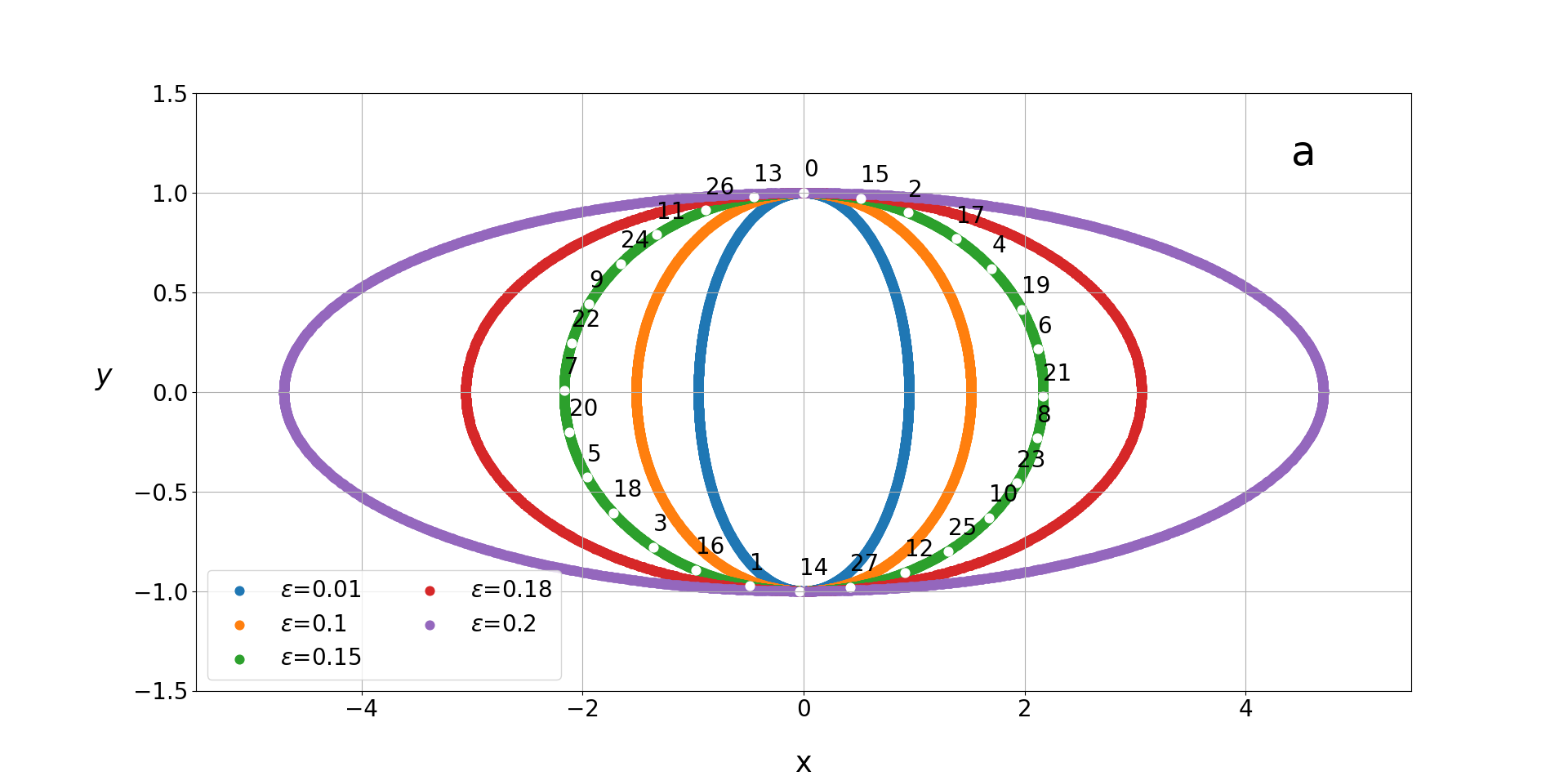}
\includegraphics[height=9cm, width=12cm]{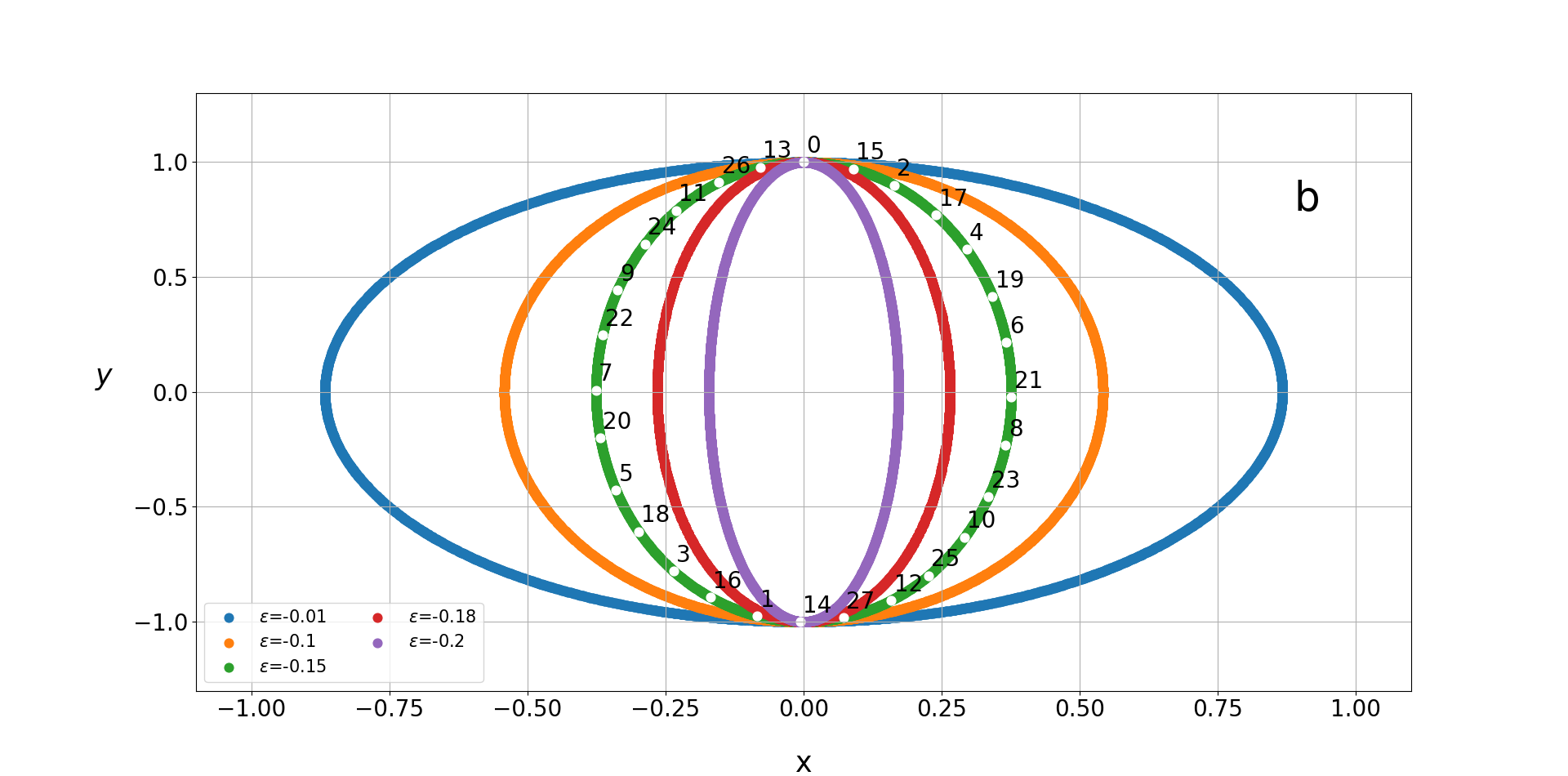}
\caption{Invariant curves on the stroboscopic surface of section in the 
case $\omega=2, \omega_1=1.1, x_0=0, y_0=1$ for various values
of $\varepsilon$ (a) for $\varepsilon>0$ the sizes of the curves increase as $\varepsilon$ increases and tend to infinity, as $\varepsilon$ tends to the critical value $\varepsilon_{crit}=0.21598$. (b) for $\varepsilon<0$ the sizes of the curves decrease as $\varepsilon$ decreases and tend to a straight line from  $y=-1$ to $y=1$ as $\varepsilon$ tends to the critical value $\varepsilon_{crit}=-0.21598$.}\label{omega_1_comma1}
\end{figure}

In all cases the critical values of $\varepsilon$ for the same
$\omega$ and $\omega_1$ are  the same for positive
and negative $\varepsilon$ but the limiting ellipse in one 
case (positive $\varepsilon$ for $\omega_1<1$, negative $\varepsilon$ for $\omega_1>1)$
  tend to the x-axis, while in the opposite
cases the ellipse tends to infinity along the x-axis.
\end{enumerate}

\section{The integrals in a resonant case}
The theory of integrals in resonant cases, independent of time, has been developed in detail
\cite{contopoulos1963resonance,Contopoulos200210}. However the applicability of the direct method of Contopoulos to resonant cases periodic in time has not yet been fully explored \cite{Contopoulos200210}.

In the present case we will construct a formal integral in a particular resonant case of the Mathieu
equation, namely in the case where the periodic term has  a frequency $\omega$ which is double of the basic
frequency $\omega_1$ of the unperturbed problem.

When $\omega_1=\omega/2=1$  the series expansion  of the integral $\Phi$ 
contains secular terms (namely time dependent terms which cannot be written in trigonometric form) and cannot be used. In  this case it is appropriate to use as zeroth order solutions  of the form
$x=\frac{\sqrt{2\Phi_0}}{\omega_1}\sin(\omega_1(t-t_0)), y=\sqrt{2\Phi_0}\cos(\omega_1(t-t_0))$ and find
\begin{align}
\nonumber\Phi_1=&\frac{1}{8}\Bigg[(y^2-x^2)\cos(2t)-2xy\sin(2t)-(y^2+x^2)\cos(2t_0)\Bigg] \\&+\frac{1}{2}(y^2+x^2)\sin(2t_0)t
\end{align}
We observe that $\Phi_1$ contains the secular term 
\begin{align}
\Phi_{1sec}=\frac{1}{2}(y^2+x^2)\sin(2t_0)t
\end{align}
However in this resonant case we can construct two more zero order integrals:
\begin{align}
\nonumber &C_0=2\Phi_0\cos(\omega t-2\omega_1(t-t_0))\stackrel{\omega=2\omega_1}{=\!=\!=\!=\!=}2\Phi_0\cos(2\omega_1 t_0)\\&
S_0=2\Phi_0\sin(\omega t-2\omega_1(t-t_0))\stackrel{\omega=2\omega_1}{=\!=\!=\!=\!=}2\Phi_0\sin(2\omega_1 t_0)
\end{align}
We start by writing $C_0$  in the form
\begin{align}
C_0=(y^2-\omega_1^2x^2)\cos(\omega t)+2\omega_1xy\sin(\omega t)
\end{align}
which is periodic in time with period $T=\frac{2\pi}{\omega}$. Then we calculate higher order terms in $C=C_0+ \varepsilon C_1+\varepsilon^2 C_2+\dots$
In particular $C_1$ is
\begin{align}
C_1&=\int_0^t\frac{\partial C_0}{\partial y}\frac{\partial H_1}{\partial x}dt
\end{align}
and after some algebra we find
\begin{align}\nonumber C_1=\frac{1}{2\omega(\omega-\omega_1)}\Big[&y^2(1-\cos(2\omega t))+x^2\Big(\cos(2\omega t)(2\omega\omega_1-\omega_1^2)+\omega_1^2-2\omega^2+2\omega\omega_1\Big)\\&-2\omega xy\sin(\omega t)\Big]
\end{align}
which does not have secular terms. Setting $\omega=2, \omega_1=1$ we have 
\begin{align}
C_1=\frac{1}{4} \Big[  \left({y}^{2} -3{x}^{2}\right)  \big( 1-\cos
 \left( 4\,t \right)  \big) -4\,xy\sin \left( 4\,t \right) \Big] 
\end{align} 
and for $t=k\pi$ we get $C_1=0$.
If we calculate $C_2$ we find
\begin{align}
C_2=\int_0^t\frac{\partial C_1}{\partial y}\frac{\partial H_1}{\partial x}dt
\end{align}
and for $\omega=2, \omega_1=1$ we find 
\begin{align}
\nonumber C_2=\frac{1}{64}&\Bigg[y^2[-10\cos(2t)+\frac{5}{3}\cos(6t)-13\cos(2t_0)+\frac{64}{3}]\\&\nonumber+x^2[-22\cos(2t)-\frac{37}{3}\cos(6t)-13\cos(2t_0)+\frac{64}{3}]\\&-2xy[6\sin(2t)-5\sin(6t)]\Bigg]-\frac{1}{8}(y^2+x^2)\sin(2t_0)t
\end{align}
Therefore $C_2$ contains the same secular term as $\Phi_1$
(up to a multiplicative constant) and
we can combine
the series of $\Phi$ and $C$ in order to avoid these terms. In particular if we multiply $\Phi$ by $\varepsilon q_1$ we find an integral 
\begin{align}
\bar{C}=C_0+\varepsilon (q_1\Phi_0+C_1)+\varepsilon^2(q_1\Phi_1+C_2)+\dots\label{Cbar}
\end{align}
that does not have any secular term up to the
order $\varepsilon^2$ if we take $q_1=\frac{1}{4}$.
In a similar way we eliminate the secular terms
of order $\varepsilon^3$ by adding an appropriate
term $\varepsilon^2q_2(\Phi_0+\varepsilon\Phi_1+\dots)$ and so on.

\begin{figure}[H]
\centering
\includegraphics[height=6.7cm, width=6.7cm]{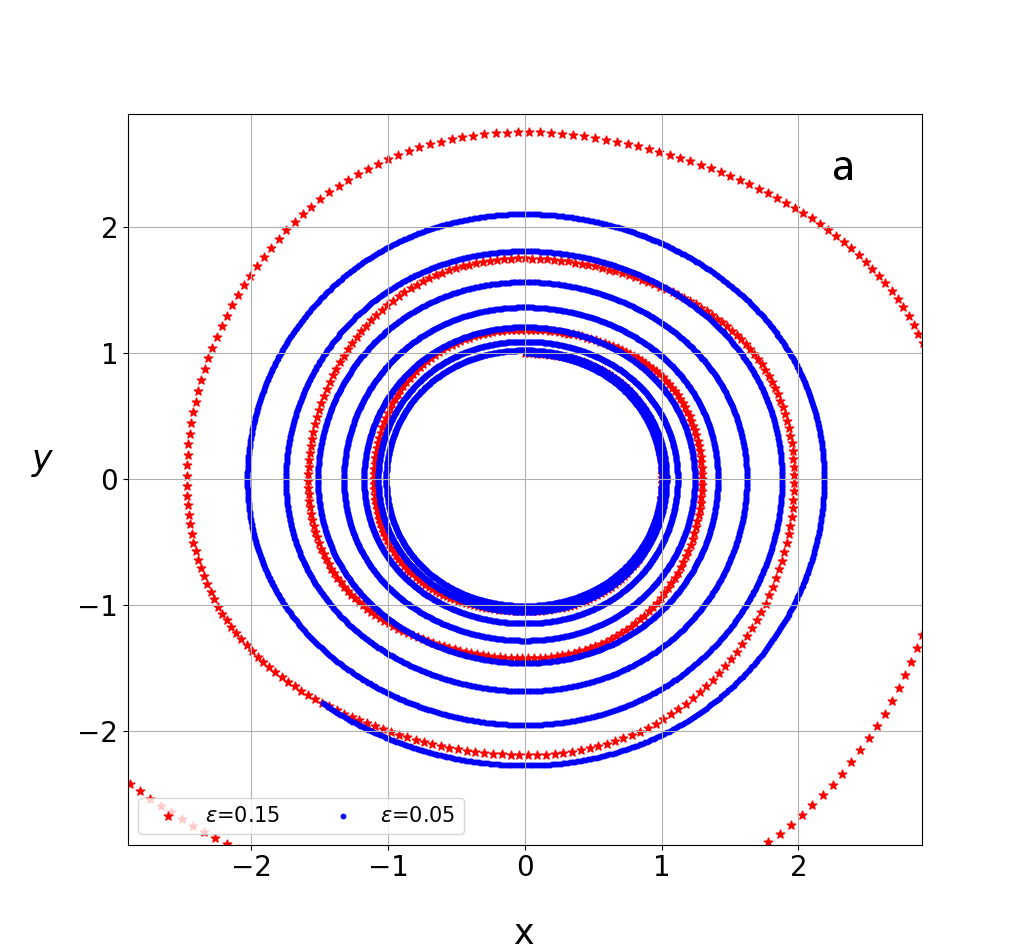}
\includegraphics[height=6.7cm, width=6.7cm]{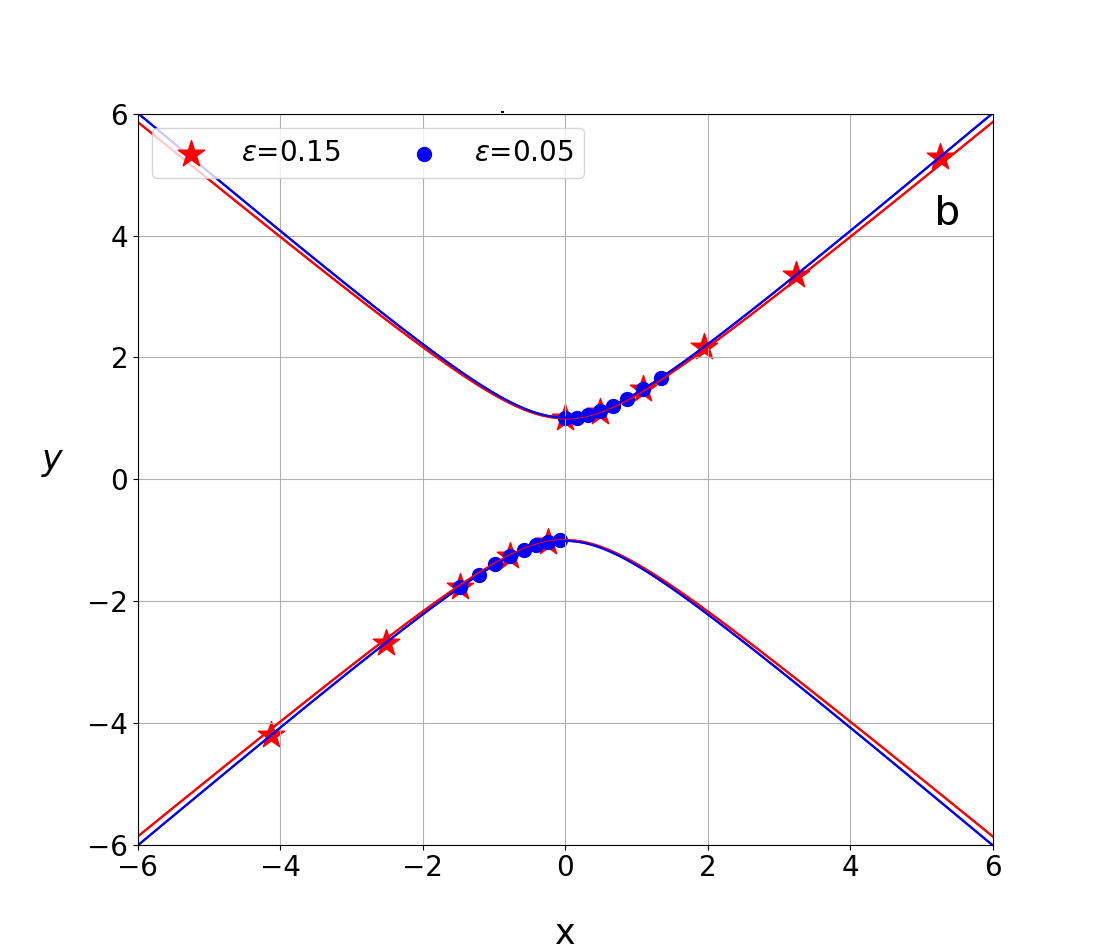}
\caption{a) Two orbits in the resonant case $\omega=2, \omega_1=1$ for  $\varepsilon=0.05$ (blue curve) and  $\varepsilon=0.15$ (red curve).
We observe the faster escape to infinity in the case of $\varepsilon=0.15$ b) The successive points (blue dots for $\varepsilon=0.05$ and red stars for $\varepsilon=0.15$) of two orbits on the stroboscopic surface of section lie, with a very good accuracy, on the correponding blue and red invariant curves given by Eq.~\eqref{Cbarst}. }\label{syntonismos}
\end{figure}

The integral \eqref{Cbar} describes a stroboscopic surface of section which for small $\varepsilon$ represents a hyperbola. Namely for $t=0$ or $t=k\pi$ we find 
\begin{align}
\bar{C}_{str}=(y^2-x^2)+\frac{\varepsilon}{4}(x^2+y^2)+\dots\label{Cbarst}
\end{align}
In Fig.~\ref{syntonismos}a we show the orbits corresponding to $\varepsilon=0.05$ and $\varepsilon=0.15$ and observe that the larger the $\varepsilon$ the faster the escape to infinity. In Fig.~\ref{syntonismos}b we draw their corresponding stroboscopic sections for times up to $t=15T$ and see that they form two hyperbolae. Then we superimpose the hyperbolae predicted by the integral \eqref{Cbarst}. We observe the very good agreement between the numerical results and the approximate integral of motion. In a similar way one may construct integrals of motion for every combination of $\omega, \omega_1$ that leads to resonances. 

\section{Conclusions}
We studied the orbits and the stroboscopic invariant curves in the Hamiltonian $H=H_0+\varepsilon H_1$, where $H_0=\frac{1}{2}(\omega_1^2x^2+y^2)$ with $y=dx/dt$ and $H_1=- x^2\cos(\omega t)$. This Hamiltonian is equivalent to a Mathieu equation. We presented a method for constructing formal integrals of motion which are written as series expansions in $\varepsilon$  and converge if $|\varepsilon|$ is smaller than a critical value $\varepsilon_{crit}$. We studied in detail the forms of the orbits and of the integrals for various values of $\varepsilon, \omega$ and $\omega_1$. Our main results are the following:
\begin{enumerate}
\item The integral is quadratic in $x$ and $y$ of the form 
\begin{align}
\Phi=C_x\omega_1^2x^2+C_yy^2+C_{xy}xy,
\end{align}
where $C_x, C_y, C_{xy}$ are series in $\varepsilon$. In non-resonant cases the integral forms invariant curves that are similar concentric ellipses on a stroboscopic Poincar\'{e} surface of section. The points of orbits with $t=kT\, (k=1,2,\dots,\, T=2\pi/\omega)$ lie on such ellipses when the series converge.
\item We found the forms of the orbits that fill, in general, elliptical rings, leaving an empty region near the center $x=y=0$. We found also the distribution of the points on the stroboscopic surface of section.
\item For particular values of $\varepsilon$ the orbits are periodic.
\item For values of $\omega$ and $\omega_1$ below the resonance $\omega-2\omega_1=0$ the stroboscopic ellipses become thinner along $x$ for $\varepsilon>0$, as $\varepsilon$ increases. As $\varepsilon\to\varepsilon_{crit}$ the ellipse tends to a straight line along the y-axis. For $\varepsilon>\varepsilon_{crit}$ the orbits spiral outwards and tend to infinity. On the other hand for $\varepsilon<0$ the ellipses become thinner along y as $\varepsilon$ decreases and as $\varepsilon\to -\varepsilon_{crit}$ they tend to a straight line along the x-axis.
\item For any given set of physical parameters $\omega, \omega_1$ and initial conditions $x_0,y_0$, there are two escape values of $\varepsilon$, one positive and one negative, with the same absolute value $|\varepsilon_{crit}|$. We checked that these values are the same as given for the escape values of $q$ by other authors \cite{mclachlan1951theory}.  On the other hand the orbits and the invariant curves are quite different for symmetric values of $\varepsilon$.

\item The present system can be considered as a 2-d Hamiltonian system with two variables $x$ and $t$ and corresponding momenta $y$ and $E$ where $-E$ is the energy. If $|\varepsilon|<\varepsilon_{crit}$ the values of $E$ oscillate between a maximum and a minimum, while if $|\varepsilon|>\varepsilon_{crit}$ the values of $|E|$ extend to infinity.
\item If $\omega-2\omega_1>0$ the orbits for $\varepsilon>0$ are of the opposite type from the orbits for $\omega-2\omega_1<0$. Namely they are more elongated along $x$ as $\varepsilon$ increases for $\varepsilon>0$ and more elongated along $y$ as $\varepsilon$ decreases for $\varepsilon<0$.
\item In resonant cases the integrals are of different form and need to be constructed for every given resonance. We made an application in one resonant case where the invariant curves are hyperbolae and are very well approximated by a second   approximate integral of motion.
\end{enumerate}
Further work will refer to more general non-linear potentials where chaos becomes apparent.

\section*{Acknowledgements}
This research was conducted in the framework of the program of the RCAAM of the Academy of Athens ``Study of the dynamical
 evolution of the entanglement and coherence in quantum systems''. The authors want to thank Dr. Christos Efthymiopoulos for his useful comments.

\bibliography{bibliography}

\end{document}